\documentclass[%
 reprint,
 amsmath,amssymb,
 aps,
]{revtex4-2}

\usepackage{graphicx}
\usepackage{dcolumn}
\usepackage{bm}

\usepackage{enumitem} 

\begin{document}


\title{ Spatial and temporal correlations in neural networks with structured connectivity}

\author{Yan-Liang Shi$^{1}$, Roxana Zeraati$^{2,3}$,  Anna Levina$^{3,4,5}$, Tatiana A. Engel$^{1}$}
\affiliation{%
$^{1}$ Cold Spring Harbor Laboratory, Cold Spring Harbor, New York, USA \\
$^{2}$ International Max Planck Research School for the Mechanisms of Mental Function and Dysfunction, University of T\"ubingen, T\"ubingen, Germany \\ 
$^{3}$ Max Planck Institute for Biological Cybernetics, T\"ubingen, Germany \\
$^{4}$ Department of Computer Science, University of T\"ubingen, T\"ubingen, Germany \\
$^{5}$ Bernstein Center for Computational Neuroscience T\"ubingen, T\"ubingen, Germany
}%

\date{\today}

\begin{abstract}
Correlated fluctuations in the activity of neural populations reflect the network's dynamics and connectivity. The temporal and spatial dimensions of neural correlations are interdependent. However, prior theoretical work mainly analyzed correlations in either spatial or temporal domains, oblivious to their interplay. We show that the network dynamics and connectivity jointly define the spatiotemporal profile of neural correlations. We derive analytical expressions for pairwise correlations in networks of binary units with spatially arranged connectivity in one and two dimensions. We find that spatial interactions among units generate multiple timescales in auto- and cross-correlations. Each timescale is associated with fluctuations at a particular spatial frequency, making a hierarchical contribution to the correlations. External inputs can modulate the correlation timescales when spatial interactions are nonlinear, and the modulation effect depends on the operating regime of network dynamics. These theoretical results open new ways to relate connectivity and dynamics in cortical networks via measurements of spatiotemporal neural correlations.
\end{abstract}

\maketitle

\section{Introduction}

Neocortical activity fluctuates endogenously on multiple spatial and temporal scales. These intrinsic fluctuations are usually quantified by correlations in neural activity.
The spatial scale of correlations is measured by equal-time cross-correlations between spike counts in pairs of neurons \cite{Cohen2011}.
The spatial correlations decrease with lateral distance between neurons in the cortex \cite{Smith2008,Smith2013,Safavi2018,Shi2022,Huang2019,Dahmen2022}.
The temporal scale of correlations is measured by the decay rate of time-delayed auto-correlation of activity in single neurons and time-delayed cross-correlations between pairs of neurons. 
Timescales of spontaneous neural activity range widely from tens of milliseconds \cite{Murray2014} up to several seconds \cite{Okun2019} and increase from sensory to association and prefrontal cortical areas \cite{Murray2014,Gao2020,Runyan2017}. 
Spatial and temporal correlations of neural activity can be modulated during changes in behavioral states, such as selective attention \cite{Harris2011,Cohen2011,Cohen2009,Mitchell2009,Ruff2016,Nandy2017,Denfield2018,Shi2022,Zeraati2021} or working memory maintenance \cite{Gao2020}, and relate to computations across different cognitive tasks \cite{Bernacchia2011,Cavanagh2016,Wasmuht2018}.
Hence, understanding how neural correlations arise from the network connectivity and dynamics will help to identify mechanisms of neural computations in the brain.

Theoretical models suggest that spatial and temporal correlations in neural activity originate from the connectivity structure of biological circuits. 
In mammalian neocortex, the wiring of neural circuits is highly structured in space. Neurons in primate cortex are organized in mini-columns which consist of ${\sim}80-100$ vertically connected neurons spanning all cortical layers \cite{Mountcastle1997,Buxhoeveden2002}. 
Minicolumns form local spatial clusters through short-range horizontal connections tiling the lateral dimension of the cortex \cite{Mountcastle1997}.
The spatial organization of local intracortical connectivity is consistent with the dependence of cross-correlations on distance \cite{Huang2019,Rosenbaum2017,Rosenbaum2017n,Darshan2018,Shi2022,Dahmen2022}.
Similarly, network models suggest that differences in timescales across cortical areas may be directly related to areal differences in recurrent connectivity strength in the primate cortex \cite{Chaudhuri2015}.

These prior theoretical studies considered either spatial or temporal dimensions of neural correlations separately. In the spatial domain, network models with spatially arranged connectivity can produce spatial patterns of neural correlations with realistic distance dependence.
This mechanism have been demonstrated in different types of network models including networks of spiking model neurons \cite{Huang2019,Rosenbaum2017,Rosenbaum2017n}, binary units \cite{Darshan2018,Zeraati2021}, and rate units \cite{Shi2022,Dahmen2022}.
In the temporal domain, theoretical studies of neural correlations focused primarily on randomly connected networks.
Recurrent interactions in these networks generate slow timescales in autocorrelation, which can be  significantly longer than the membrane time constant of individual neurons \cite{Sompolinsky1988,Hansel2015,Kadmon2015,Kumar2012,Meegen2021}. 
In these models, slow timescales can arise from operating in the transition to chaos regime \cite{Sompolinsky1988,Hansel2015,Kadmon2015} or from metastable transitions between finite randomly-connected clusters \cite{Kumar2012}.
Moreover, a heterogeneous distribution of self-coupling strengths can generate heterogeneous timescales across network units \cite{Stern2022,Chaudhuri2014}.

Temporal and spatial correlations arise from the same spatiotemporal dynamics in the network and are therefore intertwined. However, prior theoretical work did not explore the relationship between correlations in these two domains, especially in networks with spatially arranged connectivity. Theoretical understanding of how the interplay between temporal and spatial correlations arises from the network's dynamics and connectivity will provide tight constraints on models of cortical dynamics.

We show that spatial and temporal correlations are tightly interdependent in networks with stochastic dynamics and spatially arranged connectivity.
We study analytically and in numerical simulations the spatiotemporal correlations in networks of binary units with connectivity arranged in one- and two-dimensional space. 
These networks generate rich spatiotemporal patterns of activity varying across multiple temporal and spatial scales.
Using Fourier transformation, we show that each spatial frequency mode of fluctuations is related to a specific timescale, contributing hierarchically to the overall patterns of correlations.
We find that both spatial and temporal scales of correlations depend on the spatial connectivity range, and timescales associated with non-zero spatial-frequency components are heterogeneously reduced with broader spatial connectivity.
Moreover, the external input in networks with nonlinear interactions can modulate the fluctuations of mean activity and timescales of correlations with effects depending on the operating regime of network dynamics.

The organization of the paper is as follows. In Sec. II, we define the network models and derive general forms of dynamical equations for correlations. 
We use these equations to compute the spatiotemporal structure of correlations in one- (Sec. III) and two-dimensional (Sec. IV) models with spatially arranged connectivity. 
In Section V, we investigate how the input current modulates timescales of correlations in different operating regimes of network dynamics.

\section{Network models}
	
\subsection{The network architecture}
\label{network_architecture}

\begin{figure}
\includegraphics{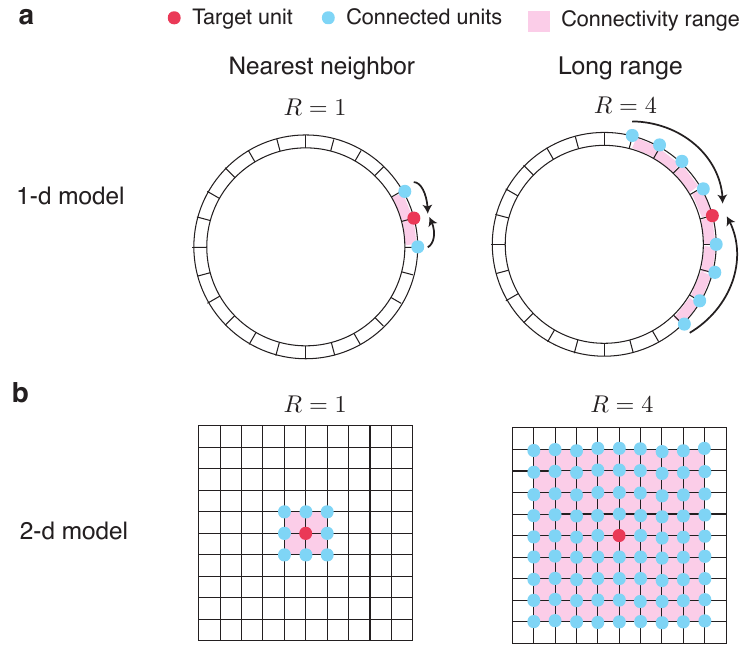}
\caption{\label{fig:schma} The network architecture.
(a) One-dimensional network with nearest neighbor ($R=1$, left) and long-range connectivity ($R>1$, right).
(b) Two-dimensional network with nearest neighbor ($R=1$, left) and long-range connectivity ($R>1$, right). 
}
\end{figure}

We consider networks of binary interacting units \cite{Ginzburg1994,Renart2010,Chow2010,Dahmen2016,Farkhooi} with spatially structured connectivity. We study one- and two-dimensional networks with different ranges of spatial connectivity.
In the one-dimensional model, $N$ units are evenly spaced on a ring with the periodic boundary condition (Fig.~\ref{fig:schma}a). In the two-dimensional model, $N^2$ units are evenly placed on the nodes of $N \times N$ square lattice with periodic boundary conditions (Fig.~\ref{fig:schma}b). In both models, units receive directed connections from their neighbors within a ball of the radius $R$ in Chebyshev distances ($L_\infty$ norm) in one or two dimensions. In models with $R=1$, each unit only receives inputs from its nearest neighbors, which we refer to as nearest neighbor connectivity. We refer to models with $R>1$ as models with long-range connectivity. 

In one-dimensional models, the perimeter of the ring is $L$, so the distance between neighboring nodes on the ring is $a=L/N$, where $a$ denotes the lattice constant. Thus, the spatial position of unit $i$ is $x_i=a \cdot i$, $i=0,...,N-1$. For connectivity radius $R$, each target unit $i$ receives directed connections from $2R$ nearby units ranging from $i-R,i-R+1,...,i+R$ ($R=1,2,...,N/2$). The strength of connectivity is uniform across these $2R$ units and scaled by $1/R$. When $R=1$, each unit $i$ only receives inputs from two nearest neighbors $i-1$ and $i+1$.

In two-dimensional models, the side length of square lattice is $L$, so both the horizontal and vertical distance between neighboring nodes are $a=L/N$, where $a$ denotes the lattice constant. We use indices ($i$, $j$) to denote a unit located at spatial position ($x_i$, $x_j$), where $x_i= i \cdot a$ and $x_j= j \cdot a$, $i,j=0,1,...,N-1$. For connectivity radius $R$, each unit ($i$, $j$) receives directed connections from $[(2R+1)^2-1]$ nearby units denoted as ($i'$, $j'$), where $i'=i-R,i-R+1,...,i+R$ and $j'=j-R,j-R+1,...,j+R$, $(i',\ j') \neq (i,\ j)$. The strength of connectivity is uniform across these $[(2R+1)^2-1]$ units and  scaled by $8/[(2R+1)^2-1]$. When $R=1$, unit ($i$, $j$) only receives inputs from eight nearest neighbors $(i',\ j')$, where $\text{max}(|i-i'|,|j-j'|)=1$.
%

\subsection{Dynamics of binary units}

In the network model, each unit $i$ can be in one of two states $S_i \in \{0, 1\}$. These states could represent the presence ($S_i =1 $) or absence ($S_i = 0 $) of a spike in a time-bin for a single neuron or the high or low activity state in a local group of neurons such as a cortical mini-column \cite{Engel2016}. 
For simplicity, we call these states active ($S_i = 1 $) and inactive ($S_i = 0 $).

The state of each unit $S_i \in \{0,1\}$ is updated based on transition rates between active-to-inactive and inactive-to-active states, given by $\omega (1 \to 0)$ and $\omega (0 \to 1)$, respectively.
We parametrize the transition rates as:
\begin{eqnarray}
\omega (0 \to 1) 	&=& \alpha_1  + \beta'_1  \  {\cal F}(\sum_{j} S_j)  \ ,  \nonumber\\
\omega (1 \to 0)    &=& \alpha_2  - \beta'_2   \ {\cal F}(\sum_{j} S_j)  \ .
\label{dynamics}
\end{eqnarray}
Here $\alpha_1$ and $\alpha_2$ are the intrinsic transition rates of one unit in the absence of network interactions. The second term in these expressions represents modulation of transition rates due to interactions with connected units in the network. In the interaction terms, $\sum_{j} S_j$ represents the sum of activity of units directly connected to unit $i$. We assume that all connected units uniformly contribute to the transition rate of the target unit (i.e. have uniform connectivity strength to the target unit). A nonlinear activation function ${\cal F}$ is a monotonically increasing function of $x$ that satisfies conditions ${\cal F}(0) =0 $, ${\cal F}(\infty) =1  $. In previous models of binary-unit networks \cite{Ginzburg1994,Renart2010,Chow2010,Dahmen2016,Farkhooi}, ${\cal F}(x) $ is usually approximated by a Heaviside function with a fixed threshold. Here we consider ${\cal F}$ of the form:
\begin{equation}
{\cal F}(\sum_{j} S_j) = 1- \exp \left(-\frac{\theta}{n} \sum_{j} S_j  \right) \ ,
\end{equation}	
where $\theta$ is a positive constant that controls the gain of recurrent inputs, and $n$ is the number of connected neighbors to each target unit. The parameters $\beta'_1$ and $\beta'_2$ control the interaction strength. To satisfy the condition of transition rate being positive, we require $\alpha_2-\beta'_2  \geqslant 0$.  In our models, we assume  
the connectivity is excitatory, hence $\beta'_1>0$ and $\beta'_2>0$. 
Thus, inputs from active neighbors will increase the transition rate from inactive to active state $\omega (0 \to 1)$ and suppress the transition rate from active to inactive state $\omega (1 \to 0)$. 
Since the connectivity is spatially organized, nearby units are more likely to become active simultaneously. Therefore, the recurrent interaction tends to enhance the spatial clustering of high activity states.
Combining $\omega (0 \to 1)$ and $\omega (1 \to 0)$, the general expression of transition rate is given by
	\begin{equation}
		\omega (S_i \to 1-S_i) = \omega (0 \to 1)  + [ \omega (1 \to 0)-\omega (0 \to 1) ] S_i   \  .
	\end{equation}

\subsubsection{Linearized approximation}
We study correlated patterns of activity fluctuations at the steady state. Outside the steady state regime, the activity is unstable and we cannot define activity fluctuations as perturbations around a fixed point. Hence, we focus on a region of model parameters in which after a sufficiently long time, the global network activity reaches an equilibrium state at a fixed point. In this case, linearization of the dynamical equations around the fixed point provides a good approximation and simplifies derivations of correlation functions. In our main analyses, we use the linear approximation of the interaction terms
\begin{equation}
		\beta'_1   \ {\cal F}(\sum_{j} S_j) \sim \beta'_1   \ {\cal F'}(0)  \cdot  \left(  \sum_{j} S_j  \right) = \beta_1  \cdot \left(   \ \sum_{j} S_j  \right) \ ,
	\end{equation}
	\begin{equation}
		\beta'_2  \ {\cal F}(\sum_{j} S_j) \sim \beta'_2  \ {\cal F'}(0)  \cdot   \left(  \sum_{j} S_j  \right) = \beta_2  \cdot \left(   \ \sum_{j} S_j  \right) \ ,
	\end{equation}
where we defined the effective interaction strengths
\begin{equation}
	\beta_1  = \beta'_1  \ {\cal F'}(0)   =  \frac{\theta}{n} \beta'_1 	 \ , \quad  \beta_2  = \beta'_2  \ {\cal F'}(0)   = \frac{\theta}{n} \beta'_2 \ .
\end{equation}
Here we assumed that the mean global activity is close to zero, so $\ {\cal F'}(S ) \approx {\cal F'}(0) $ and $\ {\cal F}( S) \approx 0$. With these conditions, the linearized transition rates become
\begin{eqnarray}
\omega (0 \to 1) 	&=& \alpha_1  + \beta_1 \cdot (\sum_{j} S_j)  \ ,  \nonumber\\
\omega (1 \to 0)    &=& \alpha_2  - \beta_2 \cdot (\sum_{j} S_j)  \ .
\label{linear_dynamics}
\end{eqnarray}
We discuss the case when mean activity $\bar S $ is non-negligible in Sect. V.

\subsubsection{Simulations of network dynamics}

\begin{figure}[b]
\includegraphics{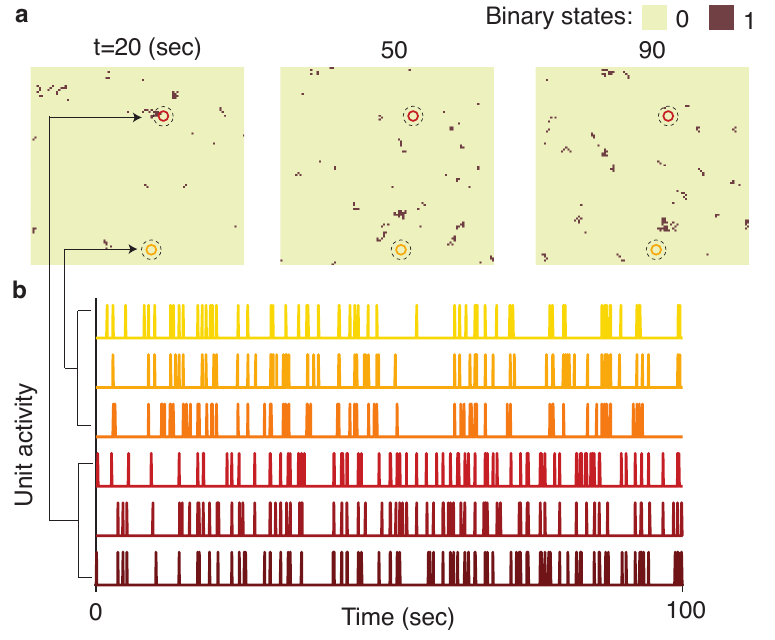}
\caption{\label{fig:simulation} Simulations of a two-dimensional network model ($N\times N=10,000$ units) with nearest-neighbor connectivity.
(a) Snapshots of population activity. 
(b) Time-series of activity states for six example units sampled from two local neighborhoods indicated with circles in a.
}
\end{figure}

We verify our analytical derivations using numerical simulations of the network models (Fig.~\ref{fig:simulation}).
We simulate the networks in discrete time using transition probabilities instead of transition rates. 
Specifically, the state of each unit is updated at each time step $t_k$ ($k$ indexes time steps with $t_{k} -t_{k-1} = \Delta t$) based on the transition probabilities:
\begin{equation}
		p (0 \to 1) = p_{\text{ext}}+  	p_r \left( \sum_{j} S_j \right) \   ,
\end{equation}
	\begin{equation}
		p (1 \to 0) = 1- p_{\text{ext}}-p_s- 	p_r \left( \sum_{j} S_j \right)  \ .
	\end{equation}
Here $p_s$ is the self-excitation probability, $p_{\textrm{ext}}$ is the probability of external excitation, and $p_r$ is the probability of recurrent excitation from active neighbors. $\sum_{j} S_j$ denotes the number of active neighbors for the target unit. This transition probability scheme is a discrete-time approximation of continuous-time dynamics with the transition rates Eq.~\ref{linear_dynamics}, linked via the parameter transformation (Appendix \ref{simulation}):
	\begin{equation}
		\alpha_1 =  p_{\text{ext}}  \left [ \frac{-\ln p_s}{(1-p_s)\Delta t} \right ]   \ ,
	    \beta_1 =  p_r \left [ \frac{-\ln p_s}{(1-p_s)\Delta t} \right ]  \ .
	\end{equation}
	\begin{equation}
		\alpha_2 =    (1-p_s-p_{\text{ext}})  \left [ \frac{-\ln p_s}{(1-p_s)\Delta t} \right ]    \ ,
	   \beta_2 =  p_r   \left [ \frac{-\ln p_s}{(1-p_s)\Delta t} \right ]   \ .
	\end{equation}

\subsection{Dynamical equations for the mean activity and correlations}

We use the master equation to derive the dynamical equation for the mean activity and the general forms of time-evolution equations for the correlation functions (Appendix~\ref{appendix3}) \cite{Ginzburg1994,Renart2010,Chow2010,Dahmen2016,Farkhooi}.
We assume $\beta_1=\beta_2$ in all calculations and model simulations unless stated otherwise.

The mean activity $\langle S_i \rangle (t)$ of unit $i$ at time $t$ (where $\langle \cdot \rangle$ denotes averaging over the distribution of all possible configurations, Eq.~\ref{average_triangle}) obeys the equation:
\begin{equation}
 \tau_0	\frac{d}{dt}\langle S_i  \rangle (t) = \frac{\alpha_1}{\alpha_1 + \alpha_2} - \langle S_i \rangle + \frac{\beta_1}{\alpha_1+\alpha_2} \langle \sum_{j } S_j \rangle    \ ,
 \label{first_moment}
\end{equation}
with $\tau_0$ is given by 
\begin{equation}
    \tau_0 = \frac{1}{\alpha_1 + \alpha_2}  \ .
    \label{equ:t0}
\end{equation}
Eq.~\ref{first_moment} shows that in the absence of  network interactions ($\beta_1=0$), the activity of each unit drifts toward the same mean value $\alpha_1/(\alpha_1+\alpha_2)$ with the intrinsic timescale $\tau_0$.
For finite interaction strength, we find the steady-state solution for the mean global activity $\bar S$ by averaging over all units $\bar S=  \sum_i \lim_{t \to\infty} \langle S_i (t)\rangle/N^{d}$, which yields
\begin{equation}
\label{bar_S}
    \bar S = \frac{\alpha_1}{\alpha_1+\alpha_2 } \cdot \frac{1}{1- n_r \left[\frac{\beta_1}{\alpha_1+\alpha_2} \right] } \ , 
\end{equation}
where $n_r$ is the number of units connected to a target unit. With the connectivity radius $R$, $n_r=2R$ for one-dimensional models, and $n_r=(2R+1)^2-1$ for two-dimensional models.
Thus, the mean activity is scaled by a factor of $1/[1-n_r \beta_1/(\alpha_1+\alpha_2)]$, which describes the effect of network interactions. 

The value of $\bar S$ sets the upper bound on the interaction strength, since $\bar S $ is a non-negative number, which implies $n_r \beta_1/(\alpha_1+\alpha_2)  < 1$. When the interaction strength exceeds this bound, the network activity becomes unstable and the mean-field approximation fails. 
We focus on the strong interaction regime which is close to the threshold of instability, i.e. $n_r \beta_1/(\alpha_1+\alpha_2) \approx 1$. For one- and two- dimensional models with nearest-neighbor connectivity, the strong interaction limit is $2\beta_1/(\alpha_1+\alpha_2) \approx 1$ and $8\beta_1/(\alpha_1+\alpha_2) \approx 1$, respectively. In Sec.~\ref{1d_CC} and \ref{2d_CC_time}, we show that in this regime, spatial recurrent interactions generate slow timescales in auto- and cross-correlations that are much longer than intrinsic timescale $\tau_0$.

To compute neural correlations, we analyze the dynamics of fluctuations around the fixed point of the mean global activity $\bar S$. We define the activity fluctuation of unit $i$ as $\delta S_i = S_i - \bar S$. The equal-time cross-correlation function is then defined as $\langle \delta S_i (t) \delta S_j (t)\rangle$ ($i \neq j$), the time-delayed cross-correlation function is $\langle \delta S_i (t) \delta S_j (t+\tau) \rangle \ (i \neq j )$, and the auto-correlation function is $\langle \delta S_i (t) \delta S_i (t+\tau) \rangle $.

The mean global activity $\bar S$ also determines the average variance of activity, which is the average auto-correlation at zero time lag: $A(0) = \sum_i \lim_{t\to \infty}\langle \delta S_i (t) \delta S_i (t) \rangle / N^d$. Using the property of binary units $\langle S_i \rangle =\langle S^2_i \rangle$, we can express $A(0)$ via $\bar S$:
\begin{eqnarray}
  A(0) &=& \lim_{t \to \infty }\frac{1}{N^d} \sum_i \langle \delta S_i (t) \delta S_i (t) \rangle \nonumber \\
 &=& \lim_{t \to \infty } \left[\frac{1}{N^d} \sum_i   \langle (S_i(t) -\bar S ) (S_i(t) -\bar S ) \rangle  \right]\nonumber\\
 & = & \lim_{t \to \infty } \frac{1}{N^d} \sum_i  \langle S^2_i \rangle - (\bar S)^2  \nonumber \\
 &=& \lim_{t \to \infty } \frac{1}{N^d} \sum_i  \langle S_i \rangle - (\bar S)^2   = \bar S (1-\bar S) \ .
\end{eqnarray}

To obtain the analytical expressions for correlations, we used the general form time-evolution equations for correlation functions (Appendix \ref{appendix3}) derived based on the master equation formalism \cite{Ginzburg1994,Renart2010,Chow2010,Dahmen2016,Farkhooi,Darshan2018}. We then applied Fourier expansion of these time-evolution equations to solve for the average equal-time and time-delayed cross-correlations and autocorrelations. Fourier expansion was used in previous work but only to study equal-time cross-correlations in one-dimensional binary network models \cite{Darshan2018} and firing-rate networks \cite{Dahmen2022} with spatial connectivity. Next, we obtained the steady-state solution based on the Fourier transformation of time-evolution of equal-time cross-correlation function. Finally, we solved the time-evolution equation of time-delayed cross-correlations and autocorrelations, where initial conditions are given by the steady state of equal-time cross-correlations. In Sec. III and IV, we  discuss the analytical solutions and numerical simulations of these correlations in different network configurations.

\section{One-dimensional models}
In this section we study spatiotemporal patterns of correlations in the one-dimensional models (Fig.~\ref{fig:schma}a).
We use Fourier transformation to derive the analytical form of the auto- and cross-correlations and study their dependence on the spatial network structure.

\subsection{Correlation functions in Fourier space}	
\label{1d_Fourier}
The one-dimensional model contains $N$ units. Each unit is located in position $x$, $x= j \cdot a,  \ j=0, ..., N-1 $, with periodic boundary condition: $ x+N\cdot a = x$ (Fig. \ref{fig:schma}a). We can expand the state $S(x)$ of the unit at position $x$ in Fourier space (here we omit the time index of $S(x)$ for notation clarity):
\begin{equation}
S(x)=  \sum_k e^{i k x} \tilde S(k) = \sum_k e^{i k j a } \tilde S(k) \ ,
\end{equation}
where $\tilde S(k)$ denotes the state variable in Fourier space with the wave number $k$. The periodic boundary condition requires the state variable to be invariant under translation $S(x+ Na) = S(x)$, which restricts the allowed values of the wave number in Fourier space: 
\begin{equation}
S(x+Na)=  \sum_k [e^{i k x} \tilde S(k)] e^{i k N a } =   \sum_k [e^{i k x} \tilde S(k)]  \ ,
\end{equation}
hence,
\begin{equation}
    e^{i k N a }= 1,  \quad    k Na  =2  \pi m, \ m= 0,  \pm 1, \pm 2, \dots .
\end{equation}
Without loss of generality, we can define the Fourier mode spectrum to be $N$ discrete values: $k=2\pi n /(Na)= 2\pi n/L$, where $n=0, 1,..., N-1 $, which is analogous to the first Brillouin zone in solid state physics \cite{Ashcroft1976}.

Similarly, we can expand the equal-time pairwise correlation function between the units located at $x_1$ and $x_2$ as (omitting the time index)
\begin{eqnarray}
C(x_1,x_2)&=& \langle \delta S(x_1) \delta S(x_2) \rangle  \nonumber \\
        &=&  \sum_{k_1} \sum_{k_2} e^{ik_1 x_1} e^{ik_2 x_2} \langle \delta \tilde S(k_1)\delta \tilde S(k_2) \rangle \ .
\end{eqnarray}
We are interested in the average correlation function $C(x_1,x_2)$ with a fixed difference between $x_1$ and $x_2$: $x=x_1-x_2$, termed $C(x)$:
\begin{equation}
C(x)=\frac{1}{N} \sum^{(N-1)a}_{x_2=0} C(x + x_2,x_2)  .
\end{equation}
This correlation function can be expanded in  Fourier space:
\begin{eqnarray}
C(x)&=& \frac{1}{N} \sum_{x_2}\langle \delta S(x+ x_2) \delta S(x_2) \rangle  \nonumber \\
        &=& \frac{1}{N}  \sum_{x_2} \sum_{k_1} \sum_{k_2} e^{ik_1 (x_2+x)} e^{ik_2 x_2} \langle \delta \tilde S(k_1)\delta \tilde S(k_2) \rangle \nonumber \\
        &=&  \sum^{\frac{2\pi(N-1)}{L}}_{k=0} e^{i k x}  \langle \delta \tilde S(k)\delta \tilde S(-k) \rangle  \nonumber \\
         &=&   \sum^{\frac{2\pi(N-1)}{L}}_{k=0} e^{i k x}  \tilde C(k) \ , 
\end{eqnarray}
where $\tilde C(k)$ is the amplitude of k-th Fourier mode of correlation function,  $\tilde C(k)=\langle \delta \tilde S(k)\delta \tilde S(-k) \rangle   = \sum_x C(x) e^{-ikx} /N $. Here, we focus on the case when the correlation function is symmetric $C(x)=C(-x)$, in which case the correlation function can be expressed as a function of distance $\Delta=|x_1-x_2|$. The distance $\Delta$ takes $N/2$ discrete values: $\Delta =n a, \ n=1, 2,...,N/2$. In this case, the Fourier modes of correlation function are restricted to take $N/2$ values: $k=0,..., (N/2-1)$:
\begin{equation}
   C(\Delta)= 2 \sum^{\frac{2\pi(N/2-1)}{L}}_{k=0} e^{i k \Delta}  \tilde C(k) \ , \quad \Delta >0  \ .
   \label{equ:c_delta}
\end{equation}
Without loss of generality, here and below we assume N to be an even integer.
By using the identity  $C(\Delta)=C(-\Delta)$, we can rewrite the Eq.~\ref{equ:c_delta} as 
\begin{equation}\label{E:CDelta}
   C(\Delta)= 2 \sum^{\frac{2\pi(N/2-1)}{L}}_{k=0} \cos( k \Delta)  \tilde C(k) \ , \quad \Delta >0  \ ,
\end{equation}
where $\tilde C(k)$ is the inverse Fourier transformation of $ C(\Delta)$: 
\begin{equation}
   \tilde C(k)= \frac{2}{N} \sum^{Na/2}_{\Delta=a} e^{-i k \Delta}   C(\Delta) \ .
\end{equation}

For the time-delayed cross-correlation function, we can also define the average correlation: 
\begin{equation}
C(x,t)= \frac{1}{N} \sum_{x_2}\langle \delta S(x+ x_2,t_0) \delta S(x_2,t_0+t)  \rangle \ ,
\end{equation}
and expand it in Fourier space as a function of distance using time-dependent Fourier amplitudes $\tilde C(k,t)$:
\begin{equation}\label{E:CDelta_t}
     C(\Delta,t)= 2 \sum^{\frac{2\pi(N/2-1)}{L}}_{k=0} \cos( k \Delta)  \tilde C(k,t) \ , \quad \Delta >0  \ .
\end{equation}
$ C(\Delta,t)$ has the initial condition $ C(\Delta,t=0) \equiv C(\Delta)$, which gives $\tilde C(k,t=0) \equiv \tilde C(k) $.

The average autocorrelation is defined as
\begin{equation}
    A(t) = \lim_{t_0 \to \infty} \sum_x \langle \delta S (x,t_0) \delta S (x,t_0+t) \rangle/ N \ .
\end{equation}
Since the average autocorrelation does not have spatial dependence, we do not directly apply Fourier transformation.

\subsection{Time-evolution equations for the average correlation functions}
We can simplify the general equations for correlations (Appendix~\ref{appendix3}) to obtain the time-evolution equations for the average correlation functions $ C(x)$,  $ C(x,t)$, $ A(t)$ defined in Sec. \ref{1d_Fourier}. By solving these equations in Fourier space, we obtain the Fourier amplitudes $\tilde C(k)$ and then compute $ C(\Delta)$, $ C(\Delta,t)$ using Eqs.~\ref{E:CDelta}, \ref{E:CDelta_t}. In the case of nearest-neighbor connectivity ($R=1$), the steady-state equation for equal-time cross-correlation function reads
\begin{eqnarray}
 C(x) &=& \frac{\beta_1}{\alpha_1+\alpha_2 }[ C(x-a) + C(x+a) \nonumber \\
& +& (\delta_{x,-a}+ \delta_{x,a})  A(0)  ] \ .
 \label{cc_r1}
\end{eqnarray}
The time evolution equation for the time-delayed cross-correlation function is 	
\begin{eqnarray}
\tau_0 \frac{d}{dt} C(x,t) = &  - & C(x,t)  \nonumber \\
& +&\frac{\beta_1}{\alpha_1+\alpha_2 } [ C(x-a,t) + C(x+a,t) \nonumber\\
&+& (\delta_{x,-a}+\delta_{x,a})  A(t)  ]  \ . 
 \label{cc_t_r1}
\end{eqnarray}		
The time evolution equation for the time-delayed auto-correlation function is 	
\begin{equation}
\tau_0 \frac{d}{dt} A(t) =  - A(t) +\frac{\beta_1}{\alpha_1+\alpha_2 }  2C(a,t)   .
 \label{ac_t_r1}
\end{equation}	
Here, we can see that the cross-correlations contribute to the auto-correlation function.

For the one-dimensional model with connectivity radius $R>1$ (i.e. long-range connectivity), we denote the average equal-time cross-correlation with fixed position difference $x$ as $C(x;R)$, the average time-delayed cross-correlation $C(x,t;R)$, and the auto-correlation $A(t;R)$. Similar to the case of nearest-neighbor connectivity, we obtain the steady-state equation for $C(x;R)$:
\begin{eqnarray}
 C(x;R) &=& \nonumber \\
\frac{\beta_1}{(\alpha_1+\alpha_2)R } & &\left [ \sum^{R}_{m=1} ( C(x-ma;R) 
 +  C(x+ma;R)) \right . \nonumber\\
 &+&  \left .    \sum^{R}_{m=1} ( \delta_{x,ma} +  \delta_{x,N-ma} ) A(0;R) \right ] \ ;
 \label{cc_rr}
\end{eqnarray}	
the time-evolution equation for $C(x,t;R)$:
\begin{eqnarray}
 &&\tau_0 \frac{d}{dt} C(x,t;R) =  - C(x,t;R)  \nonumber \\
 &&+ \frac{\beta_1}{(\alpha_1+\alpha_2)R }\left [ \sum^{R}_{m=1}(C(x-ma,t;R) + C(x+ma,t;R))  \right. \nonumber \\
 &&+ \left. \sum^{R}_{m=1} ( \delta_{x,ma} +  \delta_{x,N-ma} ) A( t ;R) \right ]   \ ;
 \label{cc_t_rr}
\end{eqnarray}		
and the time-evolution equation  for $A(t;R)$:
\begin{eqnarray}
\tau_0 \frac{d}{dt} A(t;R) & = &  - A(t;R) \nonumber \\
& + &\frac{\beta_1}{(\alpha_1+\alpha_2)R} \left [ 2\sum^{R}_{m=1}C(ma, t ;R)  \right ]  .
\label{ac_t_rr}
\end{eqnarray}

\subsection{Spatiotemporal structure of correlation functions}	
\label{1d_CC}
\subsubsection{Nearest-neighbor connectivity}

\begin{figure}[b]
\includegraphics{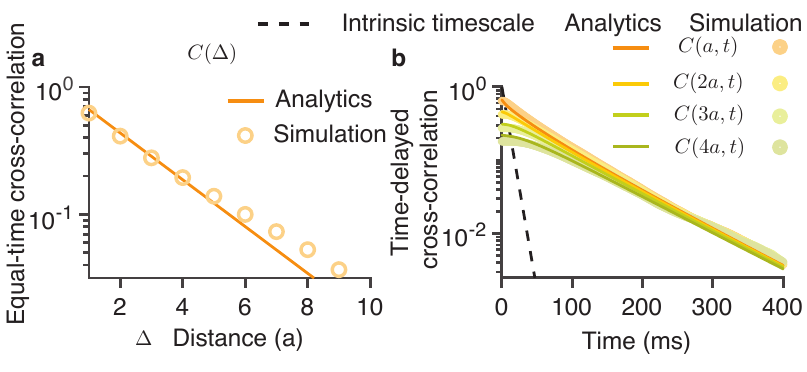}
\caption{\label{fig:cc_1d} Analytical and simulation results for the spatiotemporal cross-correlations in the one-dimensional model with nearest-neighbor connectivity. (a) Equal-time cross-correlation $C(\Delta)$ as a function of distance $\Delta$. (b) Time-delayed cross-correlation function $C(\Delta,t)$ for a range of distances ($\Delta=a,2a,3a,4a$).
The parameters are $\alpha_1=1.0653\times 10^{-4} /\Delta t$, $\alpha_2=0.1277/\Delta t$, $\beta_1=0.0586/\Delta t$, $\Delta t= 1$~ms, $N=100$.
}
\end{figure}

\begin{figure}[b]
\includegraphics{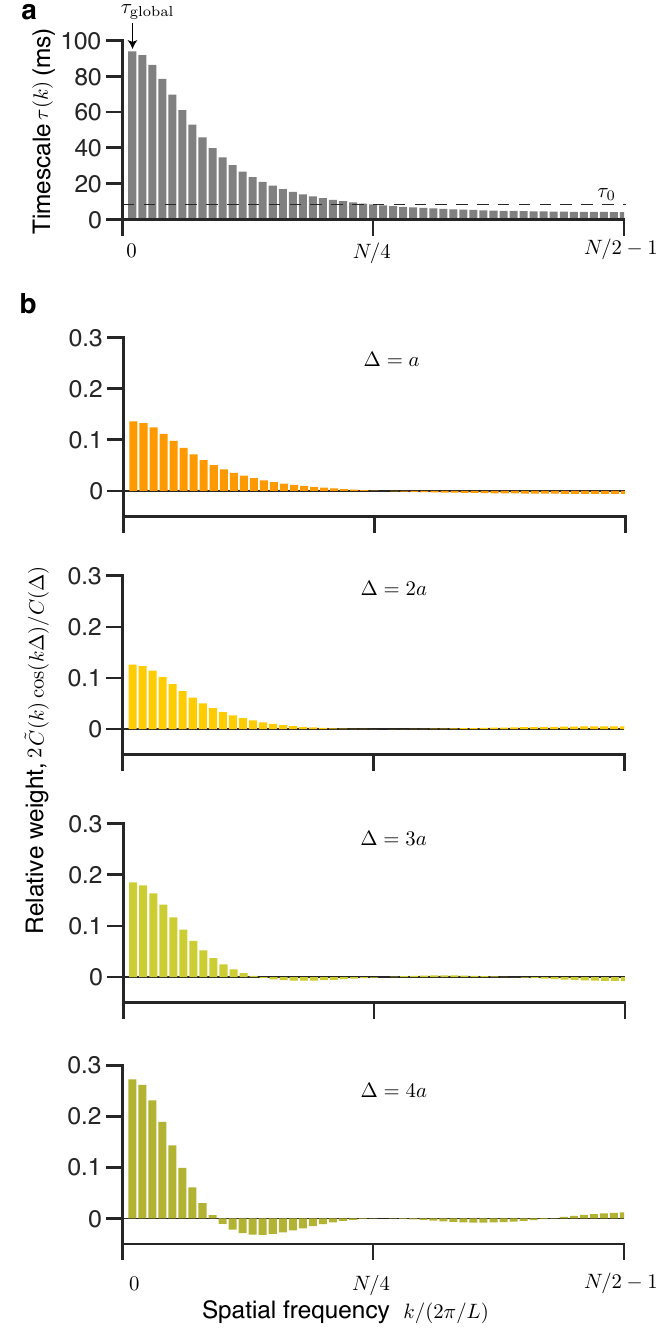}
\caption{\label{fig:cc_ck_1d} Timescales of cross-correlations $C(\Delta,t)$ and their corresponding weights for different distances $\Delta$ in the one-dimensional model with the nearest-neighbor connectivity. (a) Interaction timescales $\tau(k)$ for each spatial Fourier mode $k$. (b) Normalized weights $2\tilde C(k) \cos (k\Delta)/C(\Delta)$ for each timescale $\tau(k)$.
The parameters are $\alpha_1=1.0653\times 10^{-4} /\Delta t$,  $\alpha_2=0.1277 /\Delta t$, $\beta_1=0.0586 /\Delta t$, $\Delta t= 1$~ms, $N=100$.
}
\end{figure}

Here we study the spatiotemporal structure of correlation functions in the case of nearest-neighbor connectivity (Fig.~\ref{fig:schma}a). Eq.~\ref{ac_t_r1} describes the time evolution of autocorrelation function. The first term on the right-hand side of the equation represents the decay of autocorrelation with the rate given by the intrinsic timescale $\tau_0$. In the limit of weak interactions $\beta_1 \to 0$, we can neglect the contribution of cross-correlation to the auto-correlation and obtain the solution for autocorrelation as
\begin{equation}
    A(t)= A(0) \exp \left( - \frac{t}{\tau_0} \right) \  , \quad t \geqslant 0 .
\end{equation}
For finite interaction strength $\beta_1 >0$, the cross-correlation $C(a,t)$ acts as an external source term that brings additional temporal structures into $A(t)$. Therefore, $A(t)$ contains two types of timescales: the intrinsic timescale $\tau_0$ that is independent of network interactions, and the interaction timescales that are shared with cross-correlation $C(a,t)$.

To get the analytical form of $C(a,t)$, we first solve Eq.~\ref{cc_r1} and get $C(\Delta)$, which provides the initial condition for $C(\Delta,t)$ at $t=0$. Then, we can solve Eq.~\ref{cc_t_r1} to find $C(\Delta,t)$. Eq.~\ref{cc_r1} and Eq.~\ref{cc_t_r1} are coupled equations for $C(x)$ and $C(x\pm a)$, but they can be decoupled in Fourier space.
Using Eq.~\ref{cc_r1}, for each Fourier mode $k$, we obtain
\begin{equation}
 \tilde C(k) = \frac{2\beta_1}{\alpha_1+\alpha_2 }  \cos(ka)  \tilde C(k)  +\frac{2 \beta_1}{\alpha_1+\alpha_2 } \cos(ka)  \frac{1}{N} A(0)  \ .
\end{equation}
Then, $\tilde C(k) $ is given by
\begin{equation}
 \tilde C(k) = \frac{\frac{2 \beta_1}{\alpha_1+\alpha_2 }\cos(ka) }{1- \frac{2 \beta_1}{\alpha_1+\alpha_2 }  \cos(ka)  }  \frac{1}{N} A(0)  \ .
\end{equation}
In this expression, the factor $1/N$ comes from the normalization of the discrete Fourier transformation. The inverse Fourier transformation of $\tilde C(k)$ leads to $ C(\Delta)$:
\begin{equation}
    C(\Delta)= A(0) \exp\left(-\frac{\Delta}{L_c} \right) \ ,
\end{equation}
where we defined the correlation length $L_c$:
\begin{equation}
    L_c =a \cdot \frac{1}{\ln \left ( \frac{\alpha_1+\alpha_2}{2\beta_1}   +\sqrt{(\frac{\alpha_1+\alpha_2}{2\beta_1})^2  -1}     \right)} \ .
    \label{l_c_r1}
\end{equation}
Our analytical calculation of $C(\Delta)$ agrees well with the equal-time cross-correlation function computed from the model simulations (Fig.~\ref{fig:cc_1d}a).
In the limit of strong interactions, $2\beta_1/(\alpha_1+\alpha_2) \to 1$, the correlation length can be approximated as 
\begin{equation}
    L_c \approx a \cdot \frac{1}{\sqrt{2(\frac{\alpha_1+\alpha_2}{2\beta_1}-1)}} \ .
\end{equation}

Next, we compute the time-delayed cross-correlation.
Eq.~\ref{cc_t_r1} includes both auto- and cross-correlations. The autocorrelation $A(t)$ contains the intrinsic timescale $\tau_0$, which, as we will show, is faster than dominant timescales in the cross-correlation. Therefore, here we neglect $A(t)$ on the right-hand side of Eq.~\ref{cc_t_r1}. Under this approximation, the Fourier transformation of Eq.~\ref{cc_t_r1} is given by
\begin{eqnarray}
     \frac{d}{dt} \tilde C(k, t) & \approx & \frac{1}{\tau_0}\left[ -\tilde C(k, t) +  \frac{\beta_1}{\alpha_1+\alpha_2 }[2 \cos(ka)] \tilde C(k, t) \right] \nonumber \\
   & = & -\frac{1}{\tau(k)}  \tilde C(k, t)  .
\label{cc_k_r1}
\end{eqnarray}	
Here  $\tau(k)$ is the interaction timescale for mode $k$ (Fig. \ref{fig:cc_ck_1d}a) defined as 
\begin{equation}
    \tau(k)  = \frac{\tau_0}{1-\frac{\beta_1}{\alpha_1 + \alpha_2}[2\cos(ka)]} \ .
    \label{tau_k}
\end{equation}
Eq.~\ref{cc_k_r1} shows that each spatial Fourier mode $\tilde C(k)$ fluctuates independently with the timescale $\tau(k)$:
\begin{equation}
    \tilde C(k,t) =  \tilde C(k) \exp \left( - \frac{t}{\tau(k)} \right) \ .
\end{equation}
Thus, the time-dependence of $C(\Delta,t) $ is described by a superposition of $N/2$ Fourier modes where each mode has a characteristic timescale $\tau(k)$ with the weight $\tilde C(k)  \cos( k \Delta)$:
\begin{equation}
C(\Delta,t) =  2 \sum^{\frac{2\pi(N/2-1)}{L}}_{k=0} \tilde C(k)  \cos( k \Delta) \exp\left(-\frac{t}{\tau(k)}\right)  \ .
\label{cc_t_1d_sum_k}
\end{equation}

Our analytical calculation of $C(\Delta,t)$ agrees well with the results from numerical simulations (Fig.~\ref{fig:cc_1d}b). We find that $C(\Delta,t)$ decay in time much slower than the intrinsic timescale $\tau_0$, indicating that interaction timescales are much longer than $\tau_0$. At short time lags, the decay rate of $C(\Delta,t)$ decreases with increasing distance $\Delta$ (seen as flattening profile of $C(\Delta,t)$ at short time lags). 

To understand these distance-dependent changes in the temporal profile of correlations, we analyze the spectrum of interaction timescales $\tau(k)$ and their corresponding weights in the analytical form of $C(\Delta,t)$ (Eq.~\ref{cc_t_1d_sum_k}).
The spectrum of $\tau(k)$ is a monotonically decreasing function of $k$ in the domain $[0, 2\pi(N/2-1)/L ]$ (Fig.~\ref{fig:cc_ck_1d}a). The lowest mode, $k=0$, has the largest timescale, which we denote the global timescale:
\begin{equation}
    \tau_{\text{ global}} = \tau(k=0) =\frac{\tau_0}{1-2 \frac{\beta_1}{\alpha_1+ \alpha_2}} \ .
\end{equation}
$\tau_{\text{ global}}$ is always slower than (or equal to) the intrinsic timescale $\tau_{\text{ global}} \geqslant \tau_0$. When the interactions are very weak ($\beta_1/(\alpha_1+\alpha_2) \ll 1$), $\tau_{\text{ global}} \approx \tau_0$. In the limit of strong interactions ($1-2{\beta_1}/(\alpha_1+\alpha_2) \approx 0 $), $\tau_{\text{ global}} \gg \tau_0$. 
The corresponding spatial Fourier mode $\tilde C(k=0)= 2\sum_\Delta C(\Delta ) /N $ is the spatial average of the cross-correlation $C(\Delta)$ and describes the spatially homogeneous component of the cross-correlation.
For all other Fourier modes, timescale $\tau(k)$ decreases gradually with increasing $k$ and reaches $\tau_0$ at $k/(2\pi/L)=N/4$. For $k/ (2 \pi /L ) \in [N/4, (N/2-1) ]$, $\tau(k)$ is smaller than $\tau_0$, but weights of these modes are negligible.

To understand the structure of weights for interaction timescales $\tau(k)$ in $C(\Delta,t)$, we define the average timescale $\overline { \tau(\Delta)} $ of the correlation function 
\begin{eqnarray}
    \overline{ \tau(\Delta) }  &=& \frac{1}{C(\Delta)} \int^{+\infty}_0 C(\Delta,t) dt  =  \left [\frac{ 2\tilde C(0)}{C(\Delta)} \right] \tau_{\text{ global}}  \nonumber \\
&+& \sum^{\frac{2\pi(N/2-1)}{L}}_{k=\frac{2\pi}{L}} \left[ \frac{2\tilde C(k)  \cos( k \Delta) }{C(\Delta)} \right] \tau(k) \ .
\end{eqnarray}
This expression shows that the average timescale of cross-correlation $ \overline{\tau(\Delta)} $ is a weighted sum of $N/2$ timescales, where the normalized relative weight for mode $k$ is given by $[2\tilde C(k)  \cos( k \Delta) /C(\Delta)]$. 
These weights define the relative contribution of different Fourier modes to the cross-correlation.

The relative weights of timescales $\tau(k)$ depend on the distance $\Delta$ (Fig.~\ref{fig:cc_ck_1d}b), leading to a distance-dependent temporal profile of $C(\Delta,t)$ (Fig.~\ref{fig:cc_1d}b). In particular, the distribution of relative weights shifts towards the low-$k$ modes with increasing $\Delta$ (Fig.~\ref{fig:cc_ck_1d}b). Thus, high-$k$ modes (short-range spatial correlations) contribute to cross-correlations with larger weights at shorter distances $\Delta$, whereas low-$k$ modes (long-range spatial correlations) dominate at longer distances.
For $\Delta=a$, the relative weights monotonically decrease with $k$, with non-negligible values concentrated in the region $k/( 2 \pi /L ) \in [0, N/4 ]$ where the interaction timescales $\tau_k>\tau_0$ (Fig.~\ref{fig:cc_ck_1d}). Therefore, averaging all modes leads to $\tau_0  < \overline{ \tau(\Delta)}  < \tau_{\text{global}}$, which explains the magnitude of slope of $C(\Delta=a,t)$ (Fig. \ref{fig:cc_1d}b). 
For larger $\Delta$, the range of $k$ with non-negligible positive weights shifts toward smaller values, enhancing the relative contributions of larger timescales $\tau(k)$. As a result, $ \overline{ \tau(\Delta) }$ is positively correlated with $\Delta$. Moreover, when $\Delta>a$ (e.g., $\Delta=4a$ in Fig.~\ref{fig:cc_ck_1d}b), there are negative weights for $k$-modes in the range $k/ (2 \pi /L ) \in [0, N/4 ]$, which produce difference-of-exponentials components ($a_i\exp(-t/\tau_i)-a_j\exp(-t/\tau_j)$) in correlations. These components lead to a slow decay of correlations at short time lags, flattening the temporal profile of correlations  (Fig.~\ref{fig:cc_1d}b).

\begin{figure}
\includegraphics{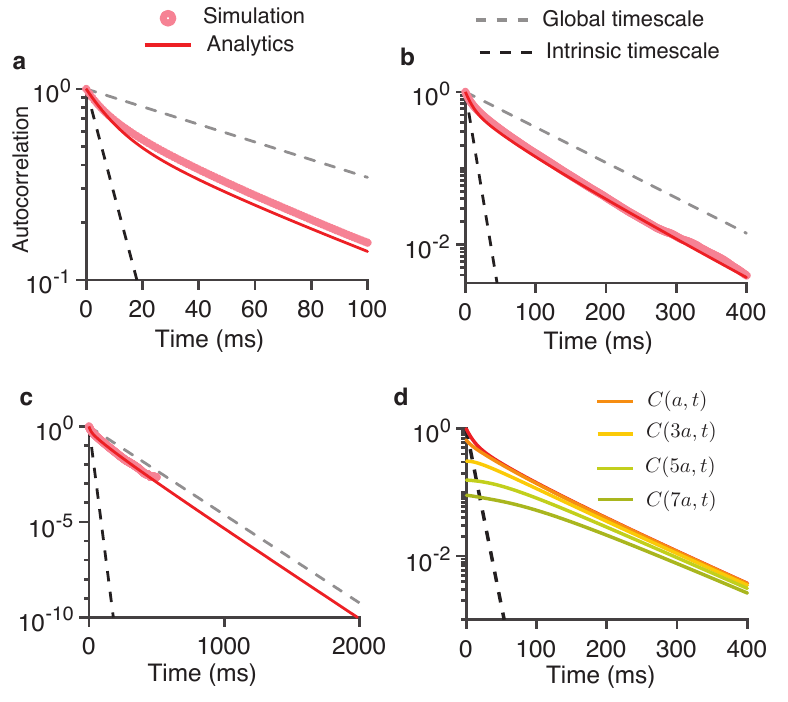}
\caption{\label{fig:AC_1d_t} Analytical and simulation results for autocorrelation $A(t)$ in the one-dimensional model with nearest-neighbor connectivity. (a)-(c) Autocorrelation function for different ranges of time lags (short - a, intermediate - b, long - c). Red line - analytical solution, pink dots - simulation results, black dashed line - exponential function with the decay rate set by the intrinsic timescale, grey dashed line - exponential function with the decay rate set by the global timescale. (d) Comparison of temporal profiles of auto- and cross-correlations. 
The parameters are $\alpha_1=1.0653\times 10^{-4}/\Delta t$,  $\alpha_2=0.1277 /\Delta t$, $\beta_1=0.0586/\Delta t$, $\Delta t= 1$~ms, $N=100$.
}
\end{figure}

Using the analytical approximation of $C(\Delta,t)$, we can solve Eq.~\ref{ac_t_r1} to obtain the analytical form of autocorrelation:
\begin{eqnarray}
    A(t) &=& A(0)\exp \left( - \frac{t}{\tau_0} \right) \nonumber \\
        &+&   2 \sum^{\frac{2\pi(N/2-1)}{L}}_{k=0} \frac{\tau(k)}{\tau(k)-\tau_0} \frac{2\beta_1}{\alpha_1+\alpha_2} \tilde C(k)  \cos( k a)  \nonumber\\
        &\times &\left[ \exp\left(-\frac{t}{\tau(k)}\right) \right]    \nonumber \\
     &=& A(0)\exp \left( - \frac{t}{\tau_0} \right) \nonumber \\
     &+&  2 \sum^{\frac{2\pi(N/2-1)}{L}}_{k=0} \tilde C(k)  \left[ \exp\left(-\frac{t}{\tau(k)}\right) \right]  \ . 
     \label{AC_1d_analytics}
\end{eqnarray}
This equation shows that $A(t)$ contains $N/2+1$ timescales: the intrinsic timescale $\tau_0$ (Eq.~\ref{equ:t0}) and $N/2$ interaction timescales $\tau(k)$ (Eq.~\ref{tau_k}) inherited from the cross-correlation. The mixture of these timescales defines the temporal profile of autocorrelation. At short time lags, the decay of autocorrelation is dominated by the intrinsic timescale (Fig.~\ref{fig:AC_1d_t}a). At intermediate time lags, the autocorrelation decays with a characteristic timescale similar to $\overline{ \tau(\Delta) }$ which is between $\tau_0$ and $\tau_{\text{global}}$ (Fig.~\ref{fig:AC_1d_t}b). In the limit of long time lags, the timescale of decay approaches the global timescale $\tau_{\text{global}}$ (Fig.~\ref{fig:AC_1d_t}c). In addition, at time lags much larger than $\tau_0$, the autocorrelation and cross-correlations decay at a similar rate (Fig.~\ref{fig:AC_1d_t}d), confirming the effects of shared Fourier amplitudes $\tilde C(k) $ in auto- and cross-correlations (Eq.~\ref{AC_1d_analytics}).

\subsubsection{Long-range connectivity}

\begin{figure}
\includegraphics{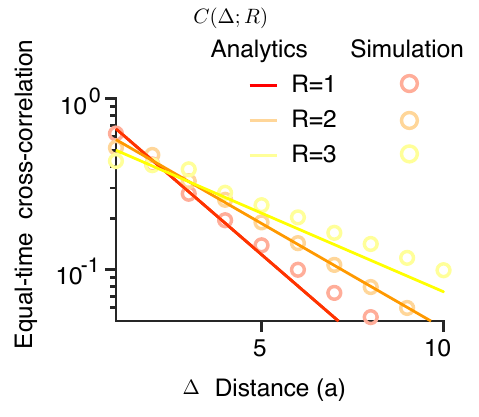}
\caption{\label{fig:CC_distance_R_1d}  Equal-time cross-correlations $C(\Delta;R)$ as a function of distance for the one-dimensional model with different connectivity radius $R$. The parameters are $\alpha_1=1.0653\times 10^{-4}/\Delta t$, $\alpha_2=0.1277/\Delta t$, $\beta_1=0.0586/\Delta t$, $\Delta t= 1$~ms, $N=100$.
}
\end{figure}

\begin{figure}[b]
\includegraphics{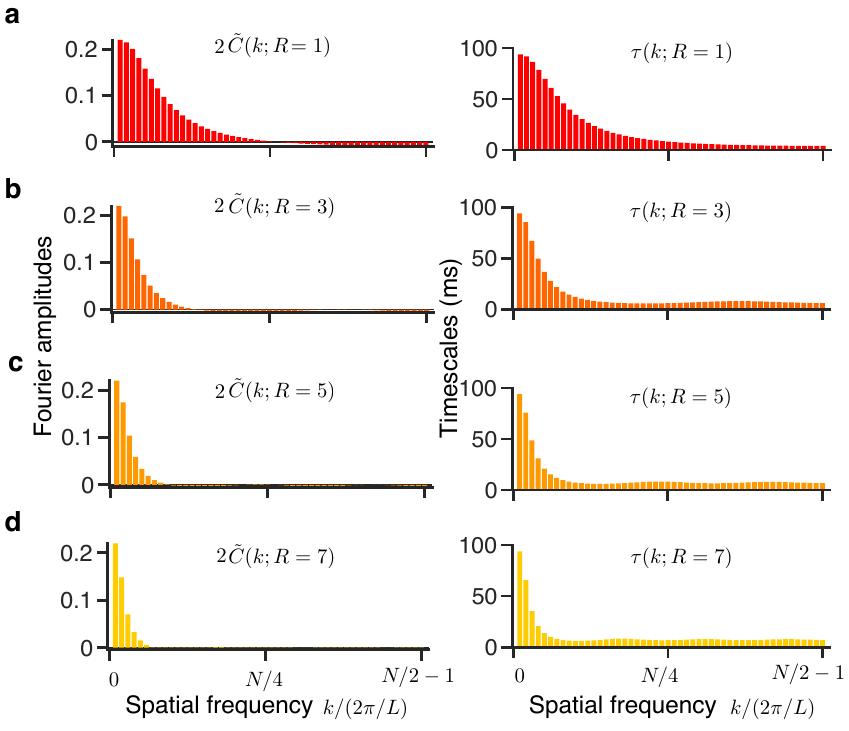}
\caption{\label{fig:AC_tau_k_1d} 
The amplitudes of spatial Fourier components $2 \tilde C(k;R)$ (left) and the corresponding interaction timescales $\tau(k;R)$ (right) of the correlation function for one-dimensional models with different connectivity radius $R$ (a, $R=1$; b, $R=3$; c, $R=5$; d, $R=7$).
The parameters are $\alpha_1=1.0653\times 10^{-4}/\Delta t$,  $\alpha_2=0.1277/\Delta t$, $\beta_1=0.0586/\Delta t$, $\Delta t= 1$~ms, $N=100$.
}
\end{figure}

Here we study correlations in one-dimensional models with long-range connectivity ($R>1$, Fig.~\ref{fig:schma}a). We investigate how the connectivity radius $R$ affects the spatiotemporal patterns of correlations. Same as for the nearest-neighbor connectivity, we solve the steady-state equation for cross-correlation Eq.~\ref{cc_rr} in Fourier space. The Fourier amplitudes of equal-time cross-correlation $\tilde C(k;R)$ are given by  
\begin{equation}
 \tilde C(k;R) = \frac{\frac{2\beta_1}{(\alpha_1+\alpha_2)R}\frac{\sin(\frac{R}{2}ka)}{\sin(\frac{1}{2}ka)}  \cos[\frac{1}{2}ka(R+1)]}{1- \frac{2\beta_1}{(\alpha_1+\alpha_2)R }\frac{\sin(\frac{R}{2}ka)}{\sin(\frac{1}{2}ka)}  \cos[\frac{1}{2}ka(R+1)] }   \frac{1}{N} A(0)  \ .
\end{equation}
This equation shows that $k=0$ mode is independent of $R$. For all other modes, the magnitude of $\tilde C(k;R)$ decreases with increasing connectivity radius $R$, especially for high-$k$ modes (short-range correlations, Fig. \ref{fig:AC_tau_k_1d}). Thus, increasing $R$ leads to more spatially homogeneous correlations (i.e. reduces the distance-dependence of correlations). This effect is evident in the position space, where equal-time cross-correlation $C(\Delta;R)$ is given by 
\begin{equation}
    C(\Delta;R) \approx A(0) \exp\left(-\frac{\Delta}{L_R} \right) 
\end{equation}
with the correlation length $L_R$
\begin{equation}\label{E:LR_1D}
    L_R= \left(\frac{R+1}{2} \right) L_c  \ .
\end{equation}
To compute $ C(\Delta;R)$, here we used the approximation $\sin(Rka/2)/[\sin(ka/2)R] \approx 1$ when $R \gg 1$ to simplify $\tilde C(k;R)$. $L_c$ is the correlation length for the model with nearest-neighbor interactions ($R=1$, Eq.~\ref{l_c_r1}). Eq.~\ref{E:LR_1D} shows that the correlation length $L_R$ is proportional to connectivity radius $R$. With increasing $R$, the network activity is more homogeneous, which is reflected in an increase of the correlation length. 
$C(\Delta,R)$ estimated from the model simulations exhibits an increase in the correlation length (measured as the slope of $C(\Delta,R)$ in the logarithmic-linear coordinates) with increasing $R$ that is in agreement with the analytical prediction (Fig.~\ref{fig:CC_distance_R_1d}).

\begin{figure}
\includegraphics{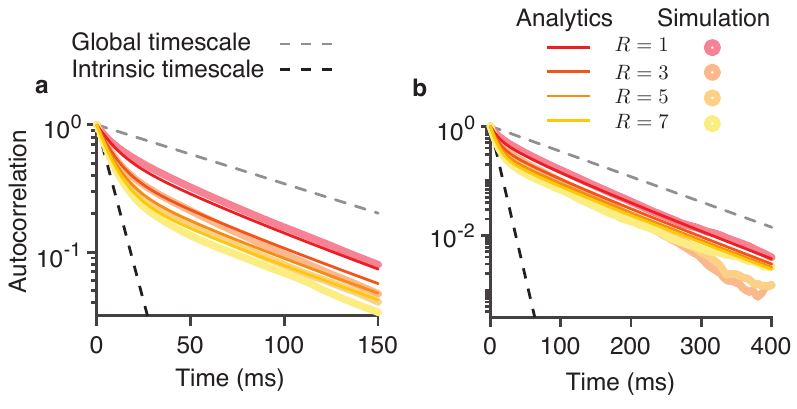}
\caption{\label{fig:AC_t_R_1d}
The autocorrelation $A(t;R) $ of the one-dimensional models with different connectivity radius $R$ plotted for different ranges of time lags (a, short; b, intermediate). Lines - analytical calculations, dots - simulation results. The parameters are $\alpha_1=1.0653\times 10^{-4}/\Delta t$,  $\alpha_2=0.1277/\Delta t$, $\beta_1=0.0586/\Delta t$, $\Delta t= 1$~ms, $N=100$.
}
\end{figure}	

To understand how the connectivity radius $R$ affects the temporal structure of correlations, we solve the equation for the time-delayed cross-correlation (Eq.~\ref{cc_t_rr}) in Fourier space, under the approximation of neglecting $A(t;R)$ (similar to the case $R=1$, Eq.~\ref{cc_k_r1}). The timescale of each mode $\tilde C(k,t;R)$ is determined by the equation:  	
\begin{equation}
    \frac{d}{dt} \tilde C(k,t;R) \approx -\frac{1}{\tau(k;R)} \tilde C(k,t;R) \ , 
\end{equation}
where the interaction timescales are
\begin{equation}
    \tau(k;R)  = \frac{\tau_0}{1- \frac{2\beta_1}{(\alpha_1+\alpha_2)R }\frac{\sin(\frac{R}{2}ka)}{\sin(\frac{1}{2}ka)} \cos[\frac{1}{2}ka(R+1)]  } \ .
    \label{tau_rr}
\end{equation} 	
Eq.~\ref{tau_rr} shows that $\tau(k;R)$ depends on the interaction radius $R$ in heterogeneous manner, depending on the value of $k$ (Fig.~\ref{fig:AC_tau_k_1d}). Specifically, for $k=0$, the global timescale is invariant to the change of $R$:
\begin{equation}
    \tau(k=0;R) \equiv \tau_{\text{global}} \ ,
\end{equation}
which means the timescale of global activity fluctuations is the same in all networks with different $R$. For finite-$k$ modes, the associated timescales decrease with increasing $R$ , and approach the intrinsic timescale $\tau_0$. In the limit of large $R$, the cross-correlation has only two non-degenerate timescales: $\tau_0$ and $\tau_{\text{global}}$. 

Next, we compute the time-delayed cross-correlation and auto-correlation functions for networks with long-range connectivity.
Summing over all Fourier modes, the time-delayed cross-correlation in the position space is given by:
\begin{equation}\label{E:CDeltatR_1D}
C(\Delta,t;R) = 2 \sum^{\frac{2\pi(N/2-1)}{L}}_{k=0} \tilde C(k;R)  \cos( k \Delta) \exp\left(-\frac{t}{\tau(k;R)}\right)   .
\end{equation}	
Combining Eq.~\ref{E:CDeltatR_1D} and Eq.~\ref{ac_t_rr}, we obtain an analytical form of autocorrelation for networks with long-range connectivity:
\begin{eqnarray}
    A(t;R) &=& A(0)\exp \left( - \frac{t}{\tau_0} \right) \nonumber \\
        &+&  2  \sum^{\frac{2\pi(N/2-1)}{L}}_{k=0}  \frac{\tau(k;R)}{\tau(k;R)-\tau_0} \frac{2\beta_1}{\alpha_1+\alpha_2} \tilde C(k;R)  \nonumber \\  &\times& \frac{1}{R}\frac{\sin(\frac{R}{2}ka)}{\sin(\frac{1}{2}ka)} \cos \left[\frac{1}{2}(R+1)ka \right] \nonumber\\
        &\times &\left[ \exp\left(-\frac{t}{\tau(k;R)}\right) \right]  \ , \nonumber \\
        &=& A(0)\exp \left( - \frac{t}{\tau_0} \right) \nonumber \\
        &+&  \sum^{\frac{2\pi(N/2-1)}{L}}_{k=0} 2\tilde C(k;R) \left[ \exp\left(-\frac{t}{\tau(k;R)}\right) \right]  .
\end{eqnarray}	

Similar to the case of nearest-neighbor interactions, $A(t;R)$ contains $N/2+1$ timescales: the intrinsic time scale $\tau_0$ (Eq.~\ref{equ:t0}) and $N/2$ interaction timescales $\tau(k;R)$ (Eq.~\ref{tau_rr}) inherited from cross-correlation $C(\Delta,t;R)$. In $A(t;R)$, the amplitude of $\tau(k;R)$ is $\tilde C(k;R)$. Changing the connectivity radius $R$ affects both the amplitudes and the corresponding timescales $\tau(k;R)$ (Fig.~\ref{fig:AC_tau_k_1d}), leading to $R$-dependent changes in autocorrelation. With  increasing $R$, the relative weight of $\tau_0$ is enhanced due to reduction of $\tilde C(k;R)$ for finite-$k$ models. Accordingly, in autocorrelation $A(t;R)$, the crossover from $\tau_0$ to the average interaction timescale (mixture of $\tau(k;R)$) occurs at a larger time lag $t$ (Fig.~\ref{fig:AC_t_R_1d}a). Since in the large time-lag limit, the autocorrelation is dominated by the largest interaction timescale  $\tau_{\text{global}}$, the autocorrelations decay at the same rate for all values of $R$ in this region (Fig.~\ref{fig:AC_t_R_1d}b).

In summary, the connectivity radius $R$ affects both the spatial and temporal structure of correlation functions. 
Increasing $R$ diminishes the non-zero spatial-frequency components in equal-time cross-correlation and suppresses the amplitude of interaction timescales (except for the global timescale) in the autocorrelation. In the large $R$ limit, cross-correlations become spatially homogeneous, and autocorrelations contain only two residual timescales, the intrinsic timescale and the global timescale. 
	
\section{Two-dimensional models}	

In this section, we generalize the analytical methods used for the one-dimensional models to study the spatiotemporal correlations in the two-dimensional models (Fig. \ref{fig:schma}b).
Similar to the one-dimensional model, we can expand the state of each unit in Fourier space.
We denote the location of units on the lattice as ($x_1$, $x_2$), where $x_1=n_1a$, $x_2=n_2a$, $n_{1,2}=0,...,N-1$. The periodic boundary conditions are $x_1 +N a= x_1$, $x_2 +N a= x_2$. 
Similar to Sect. \ref{1d_Fourier}, the periodic boundary conditions lead to discrete modes in Fourier space: $k_1=2\pi m_1/(Na)=2\pi m_1/L$, $k_2=2\pi m_2/(Na)=2\pi m_2/L$, where $m_{1,2}=0,...,N-1$. Then, the activity state of the unit at ($x_1$, $x_2$) is
\begin{equation}
    S(x_1,x_2) = \sum^{2\pi(N-1)/L}_{k_1,k_2=0} e^{ik_1 x_1} e^{ik_2 x_2} \tilde S(k_1,k_2) \ .
\end{equation}

\subsection{Correlations in two dimensions}
\begin{figure}[b]
\includegraphics{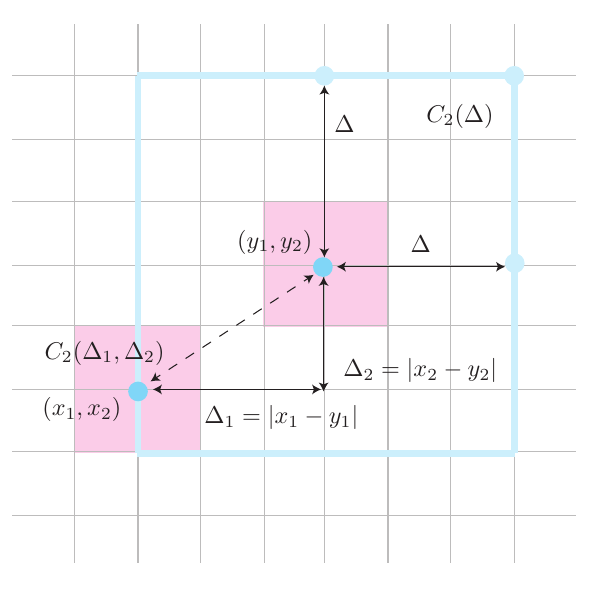}
\caption{\label{fig:4} Schematic of computing the average cross-correlation function in the two-dimensional model with nearest-neighbor interactions. Blue dots mark a pair of correlated units. Pink regions denote the range of local connectivity. The average correlation $C_2(\Delta)$ is computed by averaging correlations $C_2(\Delta_1,\Delta_2)$ over all pairs of units with $\text{max}(\Delta_1,\Delta_2)=\Delta$, which for the reference unit at $(y_1,y_2)$ corresponds to all units on the light blue square.
}
\end{figure}

The equal-time cross-correlation in two dimensions is given by 
    \begin{eqnarray}
C_2({\bf x},{\bf y})&=& \langle \delta S(x_1,x_2,t) \delta S(y_1,y_2,t) \rangle  \nonumber \\
        &=&  \sum_{k_1,k_2} \sum_{k'_1 k'_2} e^{ik_1 x_1} e^{ik_2 x_2}  e^{ik'_1 y_1} e^{ik'_2 y_2}        \nonumber \\
        &\times& \langle \delta \tilde S(k_1,k_2)\delta \tilde S(k'_1,k'_2) \rangle \ .
\end{eqnarray}
Here vectors ${\bf x}$ and  ${\bf y}$ denote $(x_1,x_2)$ and $(y_1,y_2)$, respectively.
Similar to the case of one dimension, we can define the average cross-correlation for the fixed spatial difference $x=x_1-x_2$ and $y=y_1-y_2$:
\begin{equation}
    C_2( x, y)=\frac{1}{N^2}\sum^{(N-1)a}_{x_1=0} \sum^{(N-1)a}_{y_1=0} C_2({\bf x},{\bf y})  \  .
\end{equation}
We expand $C_2(x, y)$ in Fourier space. We focus on the special case when the correlation function is symmetric, $C_2(x,y)=C_2(-x,y)$ and $C_2(x,y)=C_2(x,-y)$, and introduce $ C_2(\Delta_1, \Delta_2)$, which is the average cross-correlation for fixed two-dimensional distance ($\Delta_1$, $\Delta_2$), where $\Delta_1=|x_1-y_1|$, $\Delta_2=|x_2-y_2|$ (Fig.~\ref{fig:4}).
The distance $\Delta_{1,2}$ can take $N/2$ values, $\Delta_{1,2}= a, 2a, ...., (N/2)a$. Therefore, there are $(N/2)\times (N/2) = N^2/4$ discrete Fourier modes, $k_{1,2}=2\pi n_{1,2}/L, \ n_{1,2}=0,1,... (N/2-1)$:
\begin{equation}
   C_2(\Delta_1,\Delta_2)= 4 \sum^{\frac{2\pi(N/2-1)}{L}}_{k_1,k_2=0} \cos( k_1 \Delta_1) \cos( k_2 \Delta_2)  \tilde C_2(k_1,k_2) \ .
\end{equation}
The inverse Fourier transformation is given by 
\begin{equation}
    \tilde C_2 (k_1, k_2) = \frac{4}{N^2}\sum_{\Delta_1,\Delta_2} C( \Delta_1, \Delta_2) e^{-ik_1 \Delta_1-ik_2 \Delta_2} \ .
\end{equation}

We also define the average correlation $C_2(\Delta)$ for fixed distance $\Delta$, where we average over all pairs of $C_2(\Delta_1,\Delta_2)$ with the constraint ${\text{max}}(\Delta_1,\Delta_2)=\Delta$ (Fig.~\ref{fig:4}). $C_2(\Delta)$ can be expressed as a linear summation of $\tilde C_2 (k_1, k_2) $:
\begin{eqnarray}
     C_2(\Delta) &=&  \frac{1}{2\Delta/a}\sum_{\text{max}(\Delta_1,\Delta_2)=\Delta} C_2(\Delta_1, \Delta_2) \nonumber \\
     &=& 4 \sum^{\frac{2\pi(N/2-1)}{L}}_{k_1,k_2=0}  \frac{1}{2\Delta/a} \left [ \frac{\sin(\frac{1}{2}k_1 \Delta) }{\sin(\frac{1}{2}k_1 a)} \cos\left (\frac{1}{2}k_1(\Delta+a)\right)  \right.  \nonumber\\
      &\times& \cos(k_2\Delta) +  \left.   \frac{\sin(\frac{1}{2}k_2 \Delta) }{\sin(\frac{1}{2}k_2 a)} \cos\left (\frac{1}{2}k_2(\Delta+a)\right) \right. \nonumber \\
     &\times& \left. \cos(k_1\Delta)  \right ]  \tilde C_2(k_1,k_2)  \ .
      \label{c_2_delta}
\end{eqnarray}

The time-delayed cross-correlation in two dimensions is defined as 
\begin{equation}
    C_2({\bf x},{\bf y},t)= \langle \delta S(x_1,x_2,t_0) \delta S(y_1,y_2,t_0+t) \rangle \ .
\end{equation}
 The initial condition is given by equal-time correlations: $C_2({\bf x},{\bf y},t=0)=C_2({\bf x},{\bf y})$. Similarly, we can define average time-delayed cross-correlation $C_2( \Delta_1, \Delta_2,t)$, which has the amplitudes $\tilde C_2 (k_1, k_2, t )$ in Fourier space:
\begin{equation}
   C_2(\Delta_1,\Delta_2,t)= \sum^{\frac{2\pi(N/2-1)}{L}}_{k_1,k_2=0} \cos( k_1 \Delta_1) \cos( k_2 \Delta_2)  \tilde C_2(k_1,k_2,t) \ ,
\end{equation}
with initial condition $\tilde C_2(k_1,k_2,t=0)=\tilde C_2(k_1,k_2)$. We can also define the average correlation $C_2(\Delta,t) $ for fixed distance $\Delta$ by replacing $\tilde C_2(k_1,k_2)$ with $\tilde C_2(k_1,k_2,t)$ in Eq.~\ref{c_2_delta}.

The average autocorrelation in two-dimensional models is defined as
\begin{equation}
    A_2(t) = \lim_{t_0 \to \infty} \sum_{x,y} \langle \delta S (x,y,t_0) \delta S (x,y,t_0+t) \rangle/ N^2 \ .
\end{equation}

\subsection{Spatial structure of correlations}
\begin{figure}[b]
\includegraphics{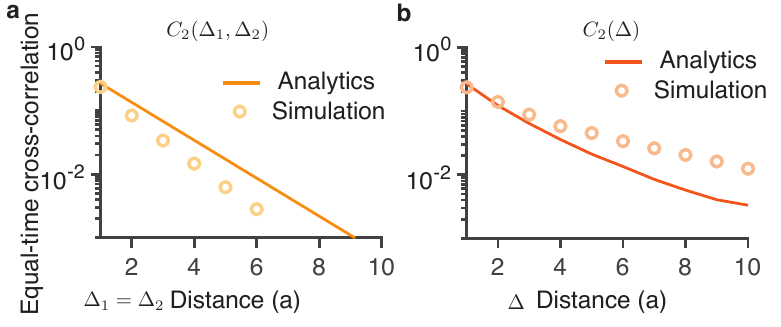}
\caption{\label{fig:d} Analytical and simulation results for the spatial dependence  of equal-time cross-correlation in the two-dimensional model with nearest-neighbor connectivity. (a) Cross-correlation $C_2(\Delta_1,\Delta_2)$ as a function of distance for $\Delta_1=\Delta_2$. (b) Average cross-correlation $C_2(\Delta)$ as a function of distance $\Delta$. 
The parameters are $\alpha_1=1.0653\times 10^{-4}/\Delta t$,  $\alpha_2=0.1277 /\Delta t$, $\beta_1=0.0146 /\Delta t$, $\Delta t= 1$~ms. Number of units: $N^2$, $N=100$.
}
\end{figure}

Next, we study the dependence of cross-correlation on the spatial distance. 
In the case of nearest-neighbor interactions, we solve the steady state equation for equal-time cross-correlation in Fourier space (Appendix \ref{2d_equatoin}) and find the amplitude of each spatial Fourier mode $(k_1,k_2)$ as
\begin{eqnarray}
 &&\tilde C_2(k_1,k_2) = \nonumber \\
 &&\frac{\frac{2\beta_1}{\alpha_1+\alpha_2 }[ 2\cos(k_1a)\cos(k_2a) + \cos(k_1a)+ \cos(k_2a)]}{1- \frac{2\beta_1}{\alpha_1+\alpha_2 }[  (2\cos(k_1a)\cos(k_2a)+\cos(k_1a)+\cos(k_2a)]  }  \nonumber \\
 && 
 \times \frac{1}{N^2} A(0)  \ . 
 \label{cc_k_2d}
\end{eqnarray}
An inverse Fourier transformation of  $\tilde C_2(k_1,k_2)$ gives rise to $ C_2(\Delta_1,\Delta_2)$:
\begin{equation}
    C_2(\Delta_1,\Delta_2)= A(0) \exp\left(-\frac{\Delta_1+\Delta_2}{L_{c,2}} \right) \ .
\end{equation}
Here $L_{c,2}$ is defined as the correlation length in two dimensions:
\begin{equation}
    L_{c,2} =a \cdot \frac{1}{\ln \left ( f   +\sqrt{f^2  -1}     \right)} \ ,
    \label{l_c_r1_2d}
\end{equation}
where
\begin{equation}
    f=-\frac{1}{2} +\frac{1}{2} \sqrt{1+\frac{\alpha_1+\alpha_2}{\beta_1}} \ .
\end{equation}
In the limit of strong interaction, $8\beta_1/(\alpha_1+\alpha_2) \to 1$, the correlation length can be approximated as 
\begin{equation}
    L_{c,2} \approx a \cdot \frac{\sqrt{3}}{2} \frac{1}{\sqrt{(\frac{\alpha_1+\alpha_2}{8\beta_1}-1)}} \ .
\end{equation}
Equal-time cross-correlations decay exponentially with increasing distance. Fig.~\ref{fig:d}a shows $C_2(\Delta_1,\Delta_2)$ as a function of distance for $\Delta_1=\Delta_2$, where $L_{c,2}$ is given by Eq.~\ref{l_c_r1_2d}.
The spatial profile of the average correlation $C_2(\Delta)$ can be obtained by combining Eqs.~\ref{c_2_delta} and \ref{cc_k_2d} (Fig. \ref{fig:d}b).

\subsection{Timescales of correlations}
\label{2d_CC_time}
Here we explore timescales of auto- and cross-correlations in the two-dimensional model with the nearest-neighbor connectivity.
\subsubsection{Cross-correlation}

\begin{figure}[b]
\includegraphics{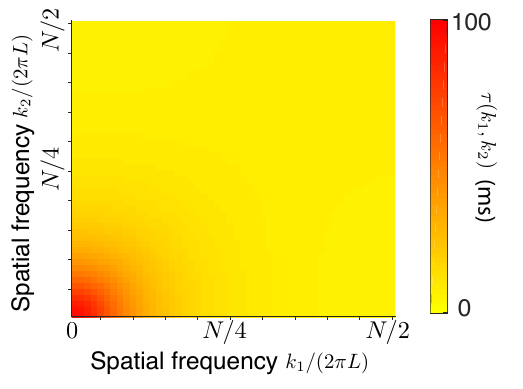}
\caption{\label{fig:tau_k} Interaction timescales $\tau(k_1,k_2)$ for Fourier modes ($k_1$,$k_2$) for the two-dimensional model with nearest-neighbor connectivity. The parameters are $\alpha_1=1.0653\times 10^{-4}/\Delta t$,  $\alpha_2=0.1277/\Delta t$, $\beta_1=0.0146/\Delta t$, $\Delta t= 1$~ms, $N=100$.
}
\end{figure}

\begin{figure}[b]
\includegraphics{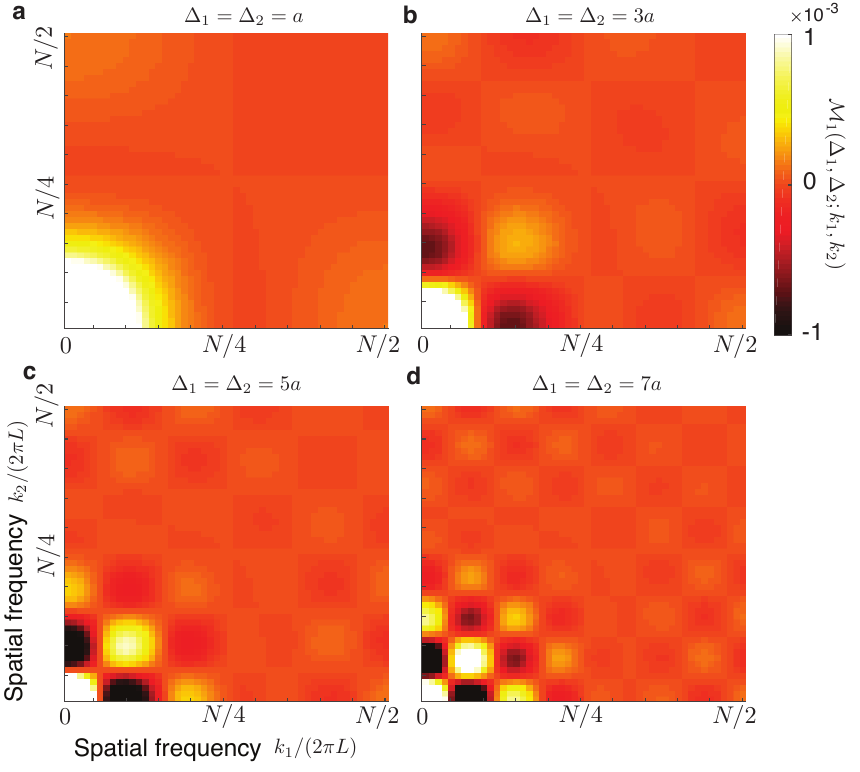}
\caption{\label{fig:M1} Weights ${\cal M}_1 (\Delta_1,\Delta_2; k_1,k_2) $ of cross-correlation $C_2(\Delta_1,\Delta_2,t)$ for each spatial frequency mode ($k_1$,$k_2$) in the two-dimensional model with nearest-neighbor connectivity. The parameters are $\alpha_1=1.0653\times 10^{-4}/\Delta t$,  $\alpha_2=0.1277/\Delta t$, $\beta_1=0.0146/\Delta t$, $\Delta t= 1$~ms, $N=100$.
}
\end{figure}

To find the temporal profiles of time-delayed auto- and cross-correlations, we first solve the time-evolution equations in Fourier space (Appendix \ref{2d_equatoin}). Under the approximation of neglecting autocorrelation in the time-evolution equation for time-delayed cross-correlations, we find that each Fourier mode $   \tilde C_2(k_1,k_2,t) $ is associated with a timescale $\tau (k_1,k_2)$ that is given by:
\begin{eqnarray}
   && \tau (k_1,k_2)  =  \nonumber \\
&&    \frac{\tau_0}{1- \frac{2\beta_1}{\alpha_1+\alpha_2 }[  2\cos(k_1a)\cos(k_2a)+\cos(k_1a)+\cos(k_2a)]  }  \ . \nonumber\\
    \label{tau_k_2d}
\end{eqnarray}
$\tau (k_1,k_2)$ is a monotonically decreasing function of $k_{1}$ and $k_{2}$ (Fig. \ref{fig:tau_k}). When $k_1=k_2=0$, $\tau (k_1,k_2)$ has the maximal value 
\begin{equation}
    \tau_{\text{global,2D}}=\tau (k_1=0,k_2=0)= \frac{\tau_0}{1-8\frac{\beta_1}{\alpha_1+\alpha_2}} \ .
\end{equation}
Analogous to the one-dimensional model, this maximal timescale $\tau_{\text{global,2D}}$ is associated with the global, spatially homogeneous mode $\tilde C(0,0)$ of fluctuations. In the limit of strong interactions ($8 \beta_1/(\alpha_1+\alpha_2) \to 1 $), the interaction timescales $\tau(k_1,k_2)  \gg \tau_0 $. At the small-$k$ region $k_{1,2}/(2\pi/L) < N/8$, the interaction timescales $\tau (k_1,k_2)$ are relatively large and $\tau (k_1,k_2) \geqslant \tau_0$. In the large $k$-region $k_{1,2}/(2\pi/L) \geqslant N/4$, the timescales  are smaller than the intrinsic timescale $\tau (k_1,k_2) \leqslant \tau_{0}$.

We can use the Fourier modes  $\tilde C_2(k_1,k_2,t)$ to describe the temporal profile of time-delayed cross-correlation $C(\Delta_1, \Delta_2,t) $.
 Each mode $   \tilde C_2(k_1,k_2,t) $ is an exponential decay function of time-lag $t$ with a time constant  $\tau (k_1,k_2)$:
\begin{equation}
    \tilde C_2(k_1,k_2,t) =  \tilde C_2(k_1,k_2) \exp \left( - \frac{t}{\tau(k_1,k_2)} \right) \ .
\end{equation}
The temporal profile of $C(\Delta_1, \Delta_2,t) $ can be described by a superposition of $N^2/4$ Fourier modes where each mode has a characteristic timescale $\tau(k_1,k_2)$ and weight $4 \tilde C(k_1,k_2)  \cos( k_1 \Delta_1)\cos( k_2 \Delta_2)$:
\begin{eqnarray}
&&C_2(\Delta_1,\Delta_2,t) = \nonumber \\
&&\sum^{\frac{2\pi(N/2-1)}{L}}_{k_1,k_2=0} 4\tilde C_2(k_1,k_2)  \cos( k_1 \Delta)  \cos( k_2 \Delta) \exp\left(-\frac{t}{\tau(k_1,k_2)}\right)  \ . \nonumber \\
&& = \sum^{\frac{2\pi(N/2-1)}{L}}_{k_1,k_2=0} {\cal M}_1( \Delta_1,\Delta_2; k_1,k_2) \exp\left(-\frac{t}{\tau(k_1,k_2)}\right)  \ .
\end{eqnarray}
Here we defined ${\cal M}_1 (\Delta_1,\Delta_2;k_1,k_2)$ to be the weight of each mode $(k_1,k_2)$:
\begin{equation}
{\cal M}_1 (\Delta_1,\Delta_2;k_1,k_2) = 4\tilde C_2(k_1,k_2)  \cos( k_1 \Delta)  \cos( k_2 \Delta) \ .
\end{equation}

\begin{figure}
\includegraphics{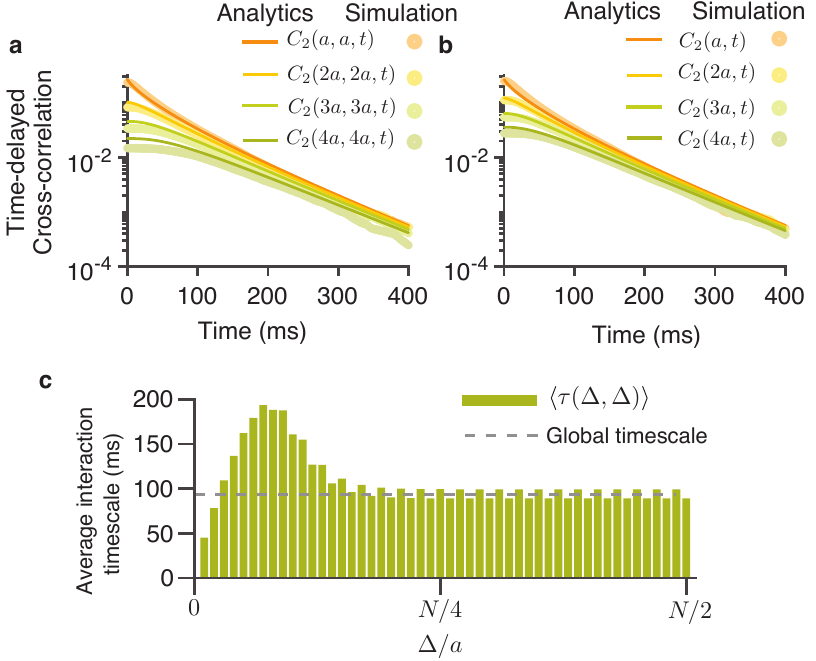}
\caption{\label{fig:cc_2d_t} Analytical and simulation results for the temporal profile of time-delayed cross-correlation in two-dimensional models with nearest-neighbor connectivity. (a) Cross-correlation function $C_2(\Delta_1,\Delta_2,t)$ for $\Delta_1=\Delta_2$ for a range of different distances $\Delta_1 $. (b) Average cross-correlation $C_2(\Delta,t)$ for a range of different distances $\Delta$. (c) Average interaction timescales of cross-correlations ($\overline{\tau (\Delta,\Delta) }$).
The parameters are $\alpha_1=1.0653\times 10^{-4}/\Delta t$, $\alpha_2=0.1277/\Delta t$, $\beta_1=0.0146/\Delta t$, $\Delta t= 1$~ms, $N=100$.
}
\end{figure}

Given the spectrum of timescales $\tau(k_1,k_2)$, the structure of ${\cal M}_1 (\Delta_1,\Delta_2;k_1,k_2)$ in $k$-space fully determines the temporal profile of cross-correlations. Since ${\cal M}_1 (\Delta_1,\Delta_2;k_1,k_2)$ also depends on the spatial distance $(\Delta_1,\Delta_2)$, the temporal and spatial scales of correlations are intertwined. For the minimal distance $\Delta_1=\Delta_2=a$, we find that the amplitude ${\cal M}_1$ is always positive in the entire domain of $k$-space (Fig.~\ref{fig:M1}a), hence  $C_2(a,a,t)$ is a monotonically decaying function of time with a super-linear slope in the logarithmic scale (Fig.~\ref{fig:cc_2d_t}a). With increasing distance $(\Delta_1,\Delta_2)$, ${\cal M}_1$ displays oscillatory patterns in $k$-space switching between positive and negative values (Fig.~\ref{fig:M1}b,c,d). In this case, some exponential components with similar timescales cancel, leading to an ultra-slow time decay (plateau) of correlations $C_2(\Delta_1,\Delta_2,t)$ at the short time lags (Fig.~\ref{fig:cc_2d_t}a).

To quantify how the average temporal profile of cross-correlations depends on distance, we compute the average interaction timescale for cross-correlation $C(\Delta_1,\Delta_2,t)$:   
\begin{eqnarray}
    &&\overline{ \tau(\Delta_1,\Delta_2) } = 
  \frac{1}{C(\Delta_1,\Delta_2)} \int^{+\infty}_0 C(\Delta_1,\Delta_2,t) dt  \nonumber \\
    &&=   \sum^{\frac{2\pi(N/2-1)}{L}}_{k_1,k_2=0} \left[ \frac{4\tilde C_2(k_1,k_2)  \cos( k_1 \Delta_1)  \cos( k_2 \Delta_2) }{C(\Delta_1,\Delta_2)} \right] \tau(k_1,k_2) \ . \nonumber\\
    \label{ave_time_CC_2d}
\end{eqnarray}
As we can see from numerical values of $\overline{ \tau(\Delta_1,\Delta_2) }$ (Fig.~\ref{fig:cc_2d_t}c), when $\Delta$ increases from the minimal distance $\Delta=a$, the average interaction timescale increases and reaches a peak at $\Delta=7a$. When $\Delta$ increases further, the average timescale decreases and approaches a value close to $\tau_{\text{global,2D}}$, because at large distances $\Delta$, $C_2(\Delta,\Delta,t)$ is dominated by the homogeneous (distance-independent) component, which has a global timescale $\tau_{\text{global,2D}}$.

\begin{figure}
\includegraphics{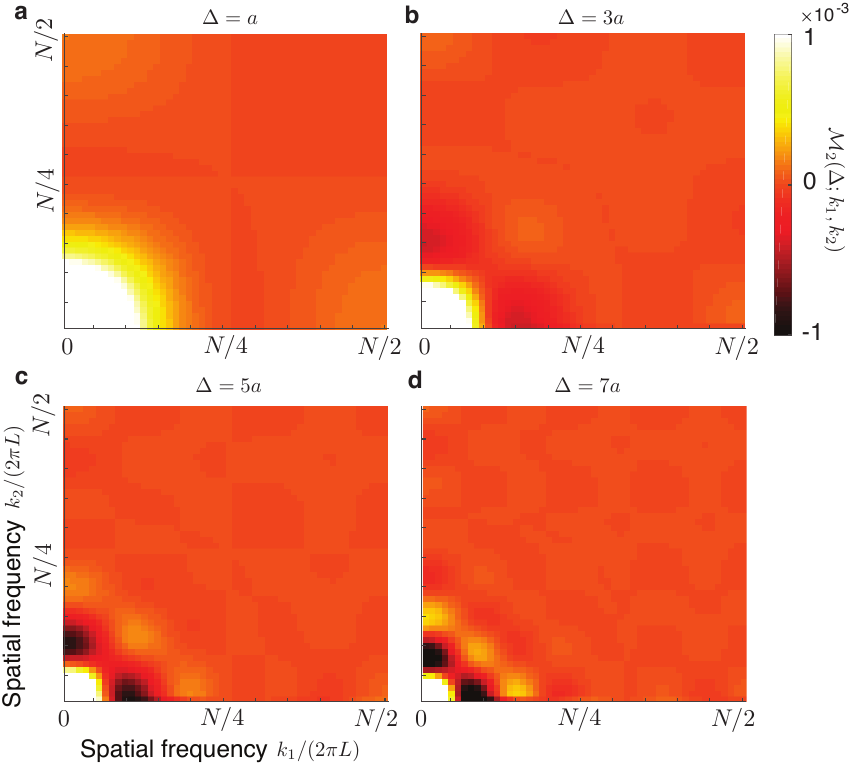}
\caption{\label{fig:M2} Weights ${\cal M}_2 (\Delta; k_1,k_2) $ of the average cross-correlation $C_2(\Delta,t)$ for each spatial frequency mode ($k_1$,$k_2$) in the two-dimensional model with nearest-neighbor connectivity. The parameters are $\alpha_1=1.0653\times 10^{-4}/\Delta t$,  $\alpha_2=0.1277/\Delta t$, $\beta_1=0.0146/\Delta t$, $\Delta t= 1$~ms, $N=100$.
}
\end{figure}

The average time-delayed correlation $C(\Delta,t) $ can also be written as a summation of  $   \tilde C_2(k_1,k_2,t) $:
\begin{eqnarray}
     C_2(\Delta,t) &=&  \frac{1}{2\Delta/a}\sum_{\text{max}(\Delta_1,\Delta_2)=\Delta} C_2(\Delta_1, \Delta_2,t ) \nonumber \\
     &=& \sum^{\frac{2\pi(N/2-1)}{L}}_{k_1,k_2=0}  \frac{4}{2\Delta/a} \left [ \frac{\sin(\frac{1}{2}k_1 \Delta) }{\sin(\frac{1}{2}k_1 a)} \cos\left (\frac{1}{2}k_1(\Delta+a)\right)  \right.  \nonumber\\
      &\times& \cos(k_2\Delta) +  \left.   \frac{\sin(\frac{1}{2}k_2 \Delta) }{\sin(\frac{1}{2}k_2 a)} \cos\left (\frac{1}{2}k_2(\Delta+a)\right)  \right. \nonumber \\
      &\times& \left. \cos(k_1\Delta)  \right ]   \tilde C_2(k_1,k_2)  \exp\left(-\frac{t}{\tau(k_1,k_2)}\right)  \nonumber \\
      & = & \sum^{\frac{2\pi(N/2-1)}{L}}_{k_1,k_2=0} {\cal M}_2( \Delta; k_1,k_2) \exp\left(-\frac{t}{\tau(k_1,k_2)}\right)  \ . \nonumber \\
\end{eqnarray}
Here we defined ${\cal M}_2 (\Delta;k_1,k_2)$ to be the weight of each mode $(k_1,k_2)$ in $C_2(\Delta,t)$. The patterns of ${\cal M}_2$ in $k$-space are shown in Fig. \ref{fig:M2} for different distances $\Delta$. Qualitatively, ${\cal M}_2$ has a similar behavior as ${\cal M}_1$. When $\Delta=a$,  ${\cal M}_2$ is always positive, creating a super-linear correlation $C_2(\Delta,t)$ in the logarithmic scale.  When $\Delta > a$, ${\cal M}_2$ oscillates between negative and positive values in the small-$k$ region, generating a slow time decay (plateau) of $C_2(\Delta,t)$ at short time lags (Fig.~\ref{fig:cc_2d_t}b).

\subsubsection{Autocorrelation}
We obtain the analytical form of autocorrelation $A_2(t)$ by solving the time-evolution equation (Appendix Eq.~\ref{ac_t_r1_2d}). In the limit of weak interactions $\beta_1/(\alpha_1+\alpha_2) \to 0 $, $A_2(t)$ is an exponential decay function with time constant $\tau_0$:  $A_2(t)= A(0) \exp(-t/\tau_0)$. For finite interaction strength, time dependence of $A_2(t)$ is influenced by the cross-correlation terms $C_2(a,a,t)$, $C_2(a,0,t)$ and  $C_2(0,a,t)$. Therefore, $A_2(t)$ inherits $N^2/4$ interaction timescales $\tau(k_1,k_2)$ (Eq.~\ref{tau_k_2d}) from the time-delayed cross-correlation. Altogether, $A_2(t)$ contains $N^2/4$ interaction timescales and an intrinsic timescale $\tau_0$, with the following analytical expression:
\begin{eqnarray}
    A_2(t) &=& A(0)\exp \left( - \frac{t}{\tau_0} \right) \nonumber \\
        &+&   4 \sum^{\frac{2\pi(N/2-1)}{L}}_{k_1,k_2=0} \frac{\tau(k_1,k_2)}{\tau(k_1,k_2)-\tau_0} \frac{2\beta_1}{\alpha_1+\alpha_2} \tilde C_2(k_1,k_2) \nonumber \\
        &\times&   [ \cos( k_1 a) +\cos( k_2 a) +2\cos( k_1 a)\cos( k_2 a)   ]\nonumber\\
        &\times &\left[ \exp\left(-\frac{t}{\tau(k_1,k_2)}\right) \right]  \nonumber \\
        &=& A(0)\exp \left( - \frac{t}{\tau_0} \right) \nonumber \\ 
        &+&  4 \sum^{\frac{2\pi(N/2-1)}{L}}_{k_1,k_2=0}    \tilde C_2(k_1,k_2)  \left[ \exp\left(-\frac{t}{\tau(k_1,k_2)}\right) \right] \ . \nonumber \\
\end{eqnarray}

\begin{figure}[b]
\includegraphics{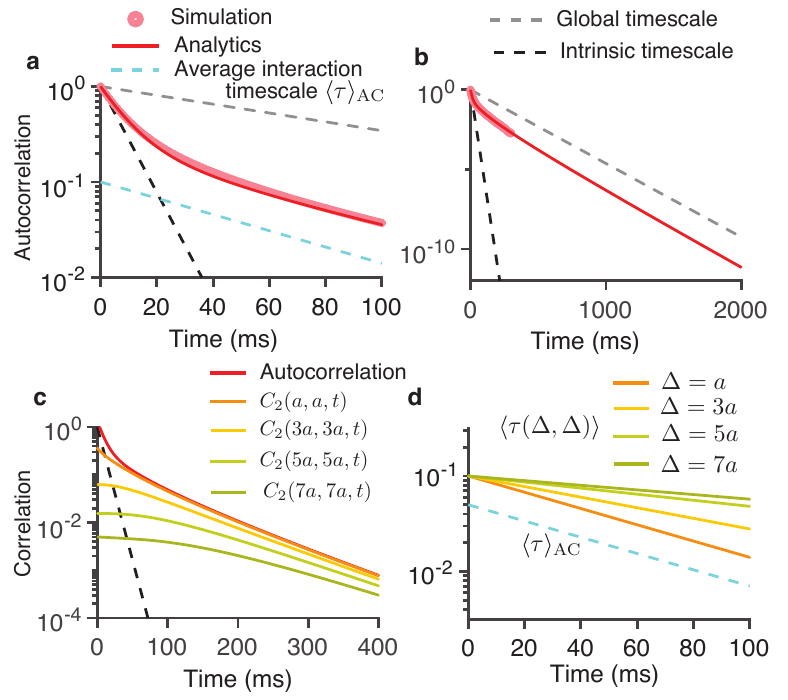}
\caption{\label{fig:AC_t} Analytical and simulation results for autocorrelation $A_2(t)$ in two-dimensional models with nearest-neighbor connectivity. (a)-(b) Autocorrelation function for different ranges of time lags (short and intermediate - a, long - b). Red line - analytical solution, pink dots - simulation results. Dashed lines - exponential functions with the decay rate set by the intrinsic timescale (black), the global timescale (grey), and the average interaction timescale $\langle \tau \rangle_{\text{AC}}$ (cyan). (c) Comparison of temporal profiles of auto- and cross-correlations. (d) Exponential functions with decay rate set by the average interaction timescales of auto- ($\langle \tau \rangle_{\text{AC}}$) and cross-correlations ($\overline{\tau (\Delta,\Delta) }$). 
The parameters are $\alpha_1=1.0653\times 10^{-4}/\Delta t$,  $\alpha_2=0.1277/\Delta t$, $\beta_1=0.0146/\Delta t$, $\Delta t= 1$~ms, $N=100$.
}
\end{figure}

The temporal decay pattern of $A_2(t)$ is dominated by different timescales at different ranges of time lags. At short time-lags, $A_2(t)$ decays with the intrinsic timescale $\tau_0$ (Fig.~\ref{fig:AC_t}a). At intermediate time lags, $A_2(t)$ decays with an intermediate timescale that is in between $\tau_0$ and $\tau_{\text{global,2D}}$  (Fig.~\ref{fig:AC_t}a). This intermediate timescale comes from a superposition of all interaction timescales $\tau(k_1,k_2)$ and is similar to the timescales of cross-correlations $C_2(\Delta_1,\Delta_2,t)$ (Fig.~\ref{fig:AC_t}c), which reflects the link between auto- and cross-correlations. In the limit of large time lags, the time decay of $A_2(t)$ is dominated by the largest timescale $\tau_{\text{global,2D}}$ and contributions of all other timescales are negligible (Fig.~\ref{fig:AC_t}b).

To quantify the average temporal profile of the interaction part of auto-correlation and its relation to the average timescales of cross-correlations, we define the average interaction timescale $\langle \tau \rangle_{\text{AC}}$ as a time integral of auto-correlation after subtracting the component associated with the intrinsic timescale:
\begin{eqnarray}
&&\langle \tau \rangle_{\text{AC}} =
\frac{\int^{+\infty}_0 [A_2(t)-A(0)\exp(-t/\tau_0)] dt  }{[A_2(t=0)-A(0)]}  \nonumber\\
&&=\sum^{\frac{2\pi(N/2-1)}{L}}_{k_1,k_2=0} \left[ \frac{4\tilde C_2(k_1,k_2)   }{C_2(0,0)} \right] \tau(k_1,k_2) \ . 
    \label{ave_time_AC_2d}
\end{eqnarray}
Here $C_2(0,0) = 4 \sum^{\frac{2\pi(N/2-1)}{L}}_{k_1,k_2=0} \tilde C_2 (k_1,k_2)  $. As we can see from numerical values of $\langle \tau \rangle_{\text{AC}}$ and  $\langle \tau(\Delta_1,\Delta_2) \rangle$ (Fig.~\ref{fig:AC_t}d), the average interaction timescale of autocorrelation is similar to the average interaction timescale of cross-correlation  $C_2(\Delta_1=a,\Delta_2=a,t)$, indicating the link between auto- and cross-correlations.

\subsection{Long-range connectivity}

\begin{figure}
\includegraphics{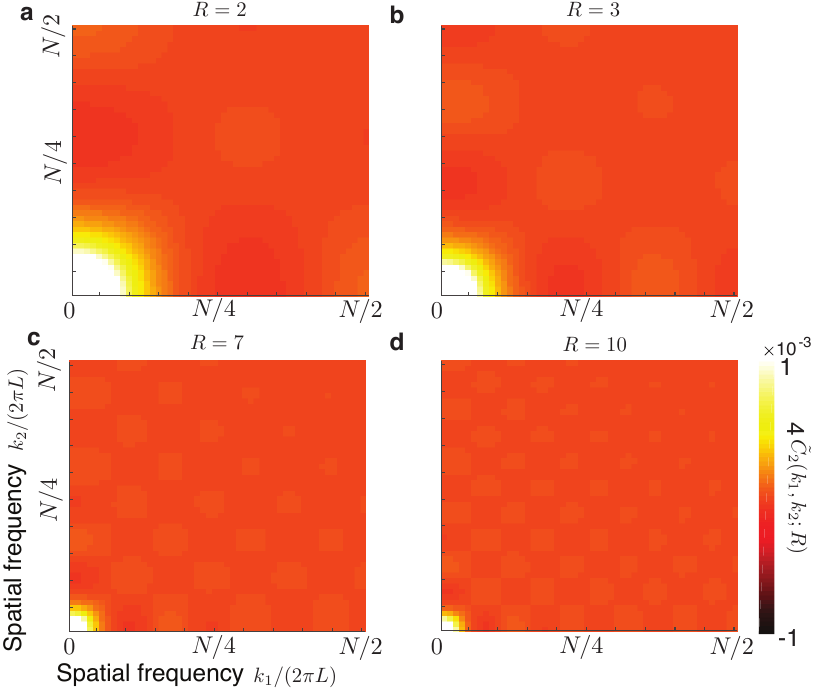}
\caption{\label{fig:C_k_R}   Fourier modes  $4\tilde C_2(k_1,k_2;R)$ for different values of connectivity radius $R$ in the two-dimensional model with long-range connectivity. The parameters are $\alpha_1=1.0653\times 10^{-4}/\Delta t$,  $\alpha_2=0.1277/\Delta t$, $\beta_1=0.0146/\Delta t$, $\Delta t= 1$~ms, $N=100$.
}
\end{figure}	

\begin{figure}
\includegraphics{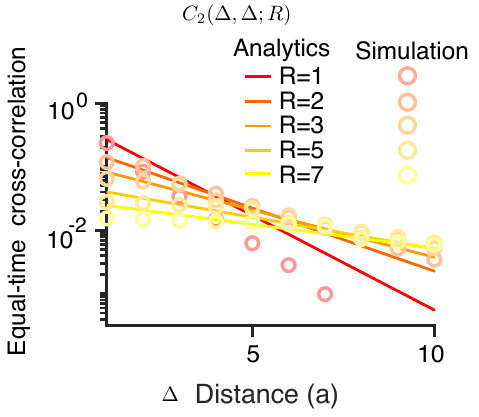}
\caption{\label{fig:CC_distance_R}  Equal-time cross-correlations $C_2(\Delta_1,\Delta_2;R)$ as a function of distance $\Delta_1=\Delta_2$ for two-dimensional models with different connectivity radius $R$. The parameters are $\alpha_1=1.0653\times 10^{-4}/\Delta t$, $\alpha_2=0.1277/\Delta t$, $\beta_1=0.0146/\Delta t$ , $\Delta t= 1$~ms, $N=100$.
}
\end{figure}		
In this section, we investigate how  the spatial extent of interactions affects the spatiotemporal correlations. We analyze correlations in two-dimensional networks with connectivity range $R>1$. In these models (Fig.~\ref{fig:schma}b), a unit $(x_1,x_2)$ connects to a unit $(y_1,y_2)$ within the range $\Delta_1,\Delta_2 \leqslant Ra$ ($\Delta_1=|x_1-y_1|,\Delta_2=|x_2-y_2|$). The strength of interactions are normalized by $8/[(2R+1)^2-1]$, such that the total recurrent input to a given unit is invariant to the change of $R$. We define equal-time cross-correlation $C_2(\Delta_1,\Delta_2;R)$, time-delayed cross-correlation $C_2(\Delta_1,\Delta_2,t;R)$, and autocorrelation $A_2(t;R)$.

Solving the time-evolution equation for the cross-correlation function with the long-range connectivity (Appendix \ref{2d_equatoin}), we find the Fourier amplitudes of equal-time cross-correlation $C_2(\Delta_1,\Delta_2;R)$: 
\begin{equation}
    \tilde C_2(k_1,k_2;R) = \frac{\frac{8\beta_1}{\alpha_1+\alpha_2 }f(k_1,k_2;R)}{1- \frac{8\beta_1}{\alpha_1+\alpha_2 }f(k_1,k_2;R)  }  \frac{1}{N^2} A(0)  \ .
 \label{cc_k_2d_rr}
\end{equation}
Here $f(k_1,k_2;R)$ is defined as 
\begin{eqnarray}
&&  f(k_1,k_2;R) =\left [\frac{1}{(2R+1)^2-1} \right] \nonumber \\
&&    \times \left[\left(1+2\frac{\sin(\frac{R}{2}k_1a)}{\sin(\frac{1}{2}k_1a)} \right. \cos(\frac{1}{2}(R+1)k_1a)\right) \nonumber \\
&& \left. \times \left(1+2\frac{\sin(\frac{R}{2}k_2a)}{\sin(\frac{1}{2}k_2a)})\cos(\frac{1}{2}(R+1)k_2a\right)-1 \right]   \ .
\end{eqnarray}
For $R\gg 1$, $f(k_1,k_2;R)$ is approximately reduced to 
\begin{eqnarray}
   && f(k_1,k_2;R) \approx  \nonumber \\
     &&  \frac{\sin(\frac{R}{2}k_1 a)\sin(\frac{R}{2}k_2 a)}{R^2\sin(\frac{1}{2}k_1a)\sin(\frac{1}{2}k_2a)} 
     \cos(\frac{1}{2}Rk_1a)\cos(\frac{1}{2}Rk_2a) \ . \nonumber \\
\end{eqnarray}
This equation shows that $f(k_1,k_2;R)$ has a maximal value at $k_{1,2}=0$ and approaches zero at $k_{1,2}=\pi/(Ra)$. Hence, the non-negligible values of $f(k_1,k_2;R)$ are restricted to the region $k_{1,2}\in [0, \pi/(Ra)]$. Therefore, $f(k_1,k_2;R)$ is a low pass filter with the width $[0, \pi/(Ra)]$. The maximal value $f(k_1=0,k_2=0;R) \equiv 1$ does not depend on $R$, whereas the band width scales with $1/R$. Hence increasing $R$ acts to reduce the number of $k$ modes that contribute to $f(k_1,k_2;R)$. The dependence of $f(k_1,k_2;R)$ on $R$ is reflected in $\tilde C(k_1,k_2;R)$. The amplitude of zero-$k$ mode $\tilde C(0,0;R)$ does not depend on $R$, and non-negligible values of $\tilde C(k_1,k_2;R)$ are restricted to the region $k_{1,2}\in [0, \pi/(Ra)]$ (Fig.~\ref{fig:C_k_R}).

As spatial scale of correlation $C_2(\Delta_1,\Delta_2;R)$ scales approximately with the inverse of Fourier wave number $k$, the correlation length should scale approximately with $R$. Indeed, we find that $ f(k_1,k_2;R) \approx  \cos((R+1) k_1 a/2)\cos((R+1) k_2 a/2)$ in the limit of $R \gg 1$, and the equal-time cross-correlation is then 
\begin{equation}
    C_2(\Delta_1,\Delta_2;R) \approx A(0) \exp\left(-\frac{\Delta_1+\Delta_2}{L_{R,2}} \right) \ ,
\end{equation}
where the correlation length is proportional to $R$:
\begin{equation}
 L_{R,2} = \left (\frac{R+1}{2}  \right ) L_{c,2} .
\end{equation}
Numerical values of $C_2(\Delta_1,\Delta_2;R)$ for intermediate $R$ confirm that the decay rate of correlations with distance (inverse of the correlation length) decreases with increasing $R$ (Fig.~\ref{fig:CC_distance_R}), which indicates the diminishing amplitudes of high wave-number modes. When $R$ reaches the maximal value  $R=N/2$, only the zero-$k$ mode has a non-zero amplitude, hence $C_2(\Delta_1,\Delta_2;R=N/2)$ becomes a homogeneous (distance independent) function. In summary, the increase of interaction-radius $R$ smooths the spatial profile of the equal-time cross-correlation by reducing the amplitudes of all non-zero wave number modes.

\begin{figure}
\includegraphics{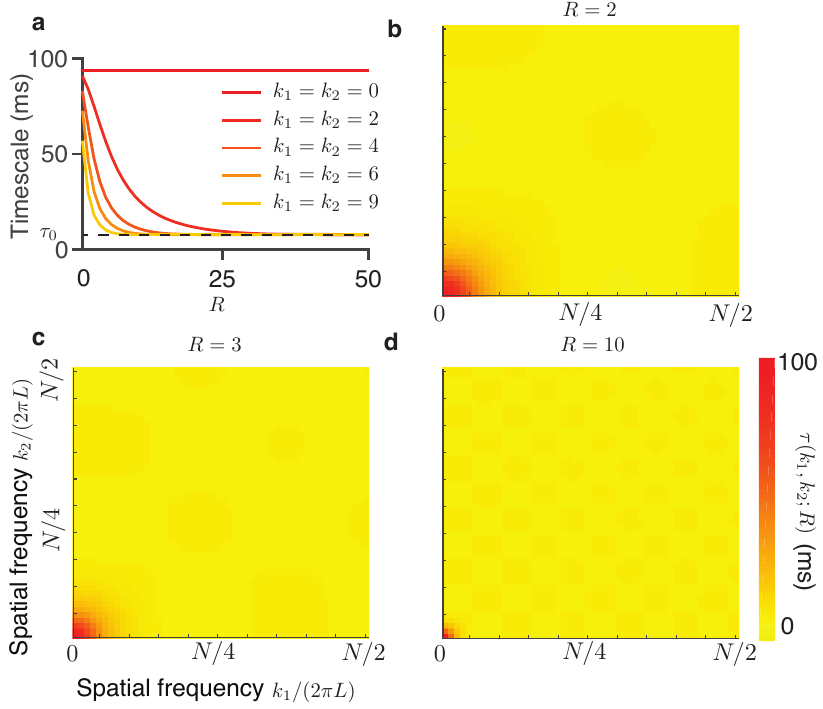}
\caption{\label{fig:tau_k_R} Interaction timescales $\tau(k_1,k_2;R)$ for the two-dimensional model with long-range connectivity. (a) $\tau(k_1,k_2;R)$ as a function of $R$ for $k_1=k_2$. (b)--(d) $\tau(k_1,k_2;R)$ in the $k$-space for different values of $R$. The parameters are $\alpha_1=1.0653\times 10^{-4}/\Delta t$, $\alpha_2=0.1277/\Delta t$, $\beta_1=0.0146/\Delta t$, $\Delta t= 1$~ms, $N=100$.
}
\end{figure}

To understand how temporal patterns of correlations depend on the connectivity range, we solve the time-evolution equation for the time-delayed cross-correlation $C_2(\Delta_1,\Delta_2,t;R)$. We solve this equation in Fourier space with the approximation of neglecting $A_2(t;R)$ terms (Appendix \ref{2d_equatoin}). We find that each mode $\tilde C_2(k_1,k_2,t;R)$ is an exponential decay function of time-lag $t$ with an interaction timescale 
\begin{equation}
 \tau (k_1,k_2;R)  =    \frac{\tau_0}{1- \frac{8\beta_1}{\alpha_1+\alpha_2 }f(k_1,k_2;R)  }  \ . 
    \label{tau_k_2d_rr}
\end{equation}
Then, the time-delayed cross-correlation can be written as a weighted sum of $N^2/4$ modes, where each mode carries an interaction timescale $ \tau (k_1,k_2;R) $:
\begin{eqnarray}
&&C_2(\Delta_1,\Delta_2,t;R) = \nonumber \\
&&4\sum^{\frac{2\pi(N/2-1)}{L}}_{k_1,k_2=0} \tilde C_2(k_1,k_2;R)  \cos( k_1 \Delta)  \cos( k_2 \Delta) \nonumber \\
&& \times \exp\left(-\frac{t}{\tau(k_1,k_2;R)}\right)  \ . 
\end{eqnarray}
Eq.~\ref{tau_k_2d_rr} shows that the magnitude of timescales $\tau (k_1,k_2;R)$ depend on $R$. In particular, the global timescale associated with $k_1=k_2=0$ mode does not depend on $R$: $\tau (k_1=0,k_2=0;R)$=$\tau_{\text{glabal,2D}}$. All other interaction timescales decrease with the increasing $R$ and are pushed towards the value of the intrinsic timescale $\tau_0$ (Fig.~\ref{fig:tau_k_R}).

\begin{figure}
\includegraphics{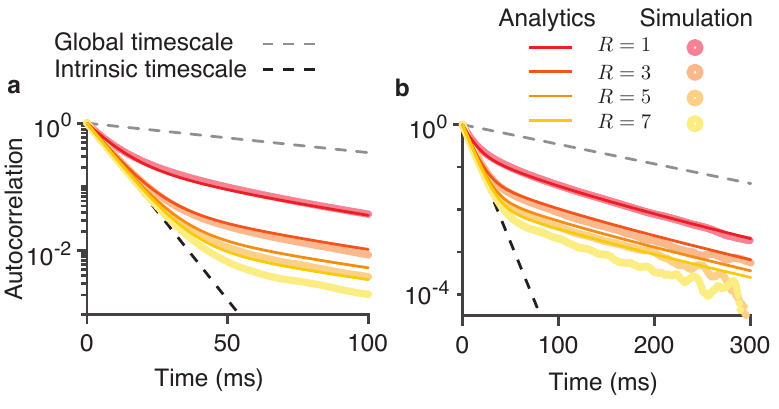}
\caption{\label{fig:AC_t_R}  Autocorrelation function $A_2(t;R) $ for different ranges of time lags (short - a, intermediate - b) for two-dimensional models with different connectivity radius $R$. Lines - analytical solution, dots - simulation results. The parameters are $\alpha_1=1.0653\times 10^{-4}/\Delta t$,  $\alpha_2=0.1277/\Delta t$, $\beta_1=0.0146/\Delta t$, $\Delta t= 1$~ms, $N=100$.
}
\end{figure}

Autocorrelation $A_2(t;R)$ is given by the combination of a component with the intrinsic timescale $\tau_0$ and $N^2/4$ components inherited from the cross-correlation modes with interaction timescales $\tau(k_1,k_2;R)$:
\begin{eqnarray}
   && A_2(t;R) = A(0)\exp \left( - \frac{t}{\tau_0} \right) \nonumber \\
    &&    +  4  \sum^{\frac{2\pi(N/2-1)}{L}}_{k_1,k_2=0} \frac{\tau(k_1,k_2;R)}{\tau(k_1,k_2;R)-\tau_0} \frac{8\beta_1}{\alpha_1+\alpha_2} \tilde C_2(k_1,k_2;R) \nonumber \\
        &&\times  f(k_1,k_2;R) \left[ \exp\left(-\frac{t}{\tau(k_1,k_2;R)}\right) \right]  \nonumber \\
        &&= A(0)\exp \left( - \frac{t}{\tau_0} \right) \nonumber \\
         &&+   4 \sum^{\frac{2\pi(N/2-1)}{L}}_{k_1,k_2=0}\tilde C_2(k_1,k_2;R)    \left[ \exp\left(-\frac{t}{\tau(k_1,k_2;R)}\right) \right] \ .   \nonumber\\
\end{eqnarray}
The temporal profile of autocorrelation is influenced by the relative weights of intrinsic timescale $\tau_0$ and the interaction timescales. At short time lags $t \approx \tau_0$, $A_2(t;R)$ decays with the timescale $\tau_0$. At intermediate time lags, $A_2(t;R)$ decays with an intermediate timescale which reflects the cumulative effect of all interaction timescales. In between these two regions, the autocorrelation slope (in the logarithmic-linear coordinates) changes abruptly indicating a crossover from decay rate dominated by $\tau_0$ to the intermediate timescales. 
Since the amplitudes $\tilde C_2(k_1,k_2;R)$ of interaction timescales decrease with increasing $R$ (except the zero-$k$ mode), the time lag where the crossover occurs increases monotonically with $R$ (Fig.~\ref{fig:AC_t_R}a). 
At large time lags, the overall decay rate is governed by $\tau_{\text{global,2D}}$. %
Since the amplitude of zero-$k$ mode with the global timescale $\tau_{\text{global,2D}}$ is independent of $R$ (Fig.~\ref{fig:C_k_R}), autocorrelations of models with different $R$ exhibit the same slope at large time lags, with different intercepts reflecting the $R$-dependence of components associated with other interaction timescales (Fig.~\ref{fig:AC_t_R}b).

	\section{Operating regime of network dynamics and timescales of correlations}
	
In previous sections, we focused on the case where the mean global activity $\bar S$ was very close to zero. Here we discuss how the mean global activity affects correlations. We show that increasing mean global activity can increase or decrease the intrinsic timescale of correlations, depending on the sign of $\beta'_1-\beta'_2$. Previous studies of binary neuron models \cite{Ginzburg1994,Renart2010,Chow2010,Dahmen2016,Farkhooi} analyzed only the special case $\beta'_1=\beta'_2$, in which the intrinsic timescale does not depend on the mean global activity. The mean global activity also affects effective interaction strengths $\beta_{1,2}$ and therefore influences the interaction timescales of correlations. We also show how the external input affects the magnitude of the mean global activity, its stability, and timescales of correlations in different operating regimes of network dynamics.

	\subsection{ The mean global activity and the intrinsic timescale of correlation }
	
The mean global activity modulates the transition rates and therefore can affect the intrinsic timescale.	
In the derivation of transition rates $\omega (0 \to 1) $ and $\omega (1 \to 0) $ (Eq.~\ref{dynamics}), we  can perform Taylor expansion around the mean global activity $\bar S$. The expansion for the interaction terms are given by 
	
	\begin{eqnarray}
		\beta'_1   \ {\cal F}(\sum_{j} S_j) &=& \beta'_1   \ {\cal F'}(n\bar S )  \   \left(  \sum_{j} S_j  \right) +  \beta'_1 {\cal F}_0      \nonumber\\
		&=& \beta_1   \left(   \ \sum_{j} S_j  \right) + \beta'_1 {\cal F}_0 \ ,
	\end{eqnarray}
	\begin{eqnarray}
		\beta'_2  \ {\cal F}(\sum_{j} S_j) &=& \beta'_2  \ {\cal F'}(n\bar S)  \   \left(  \sum_{j} S_j  \right) + \beta'_2  {\cal F}_0 
		\nonumber\\
		&=& \beta_2   \left(   \ \sum_{j} S_j  \right) +  \beta'_2  {\cal F}_0 \ ,
	\end{eqnarray}
where ${\cal F}_0 $ is defined as 
	\begin{equation}
		{\cal F}_0 = {\cal F}(n\bar S) - n \bar S {\cal  F}' (n \bar S ) + O([ (  \sum_{j} S_j  ) -n \bar S]^2) \ .
	\end{equation}
Here ${\cal F'}$ denotes the derivative of ${\cal F}$.  
Since the activation function is
\begin{equation}
{\cal F} (n\bar S)  = 1- \exp(-\theta  \bar S) \ ,
\end{equation}
 the explicit forms of ${\cal F}_0 $ and  $d {\cal F}_0 / d \bar S	 $ are
		\begin{equation}
	{\cal F}_0  \approx [1-(1+\theta \bar S)\exp(-\theta \bar S)] >0\ , \quad  0\leqslant \bar S  \leqslant 1 ;
	\end{equation}
	\begin{equation}
\frac{d}{d \bar S}	{\cal F}_0 = \theta^2 \bar S  \exp(-\theta \bar S)  > 0 \ , \quad  0\leqslant \bar S  \leqslant 1 . 
	\end{equation}
The interaction strengths in the linearized approximation are 
	\begin{equation}
	\beta_1 =  \beta'_1   \ {\cal F'}(n\bar S )  \  , \quad   \beta_2 =  \beta'_2  \ {\cal F'}(n\bar S)  \  .
	\end{equation}
When the  mean global activity $\bar S \ll 1$, we can neglect ${\cal F}_0$ term and replace $ \ {\cal F'}(n\bar S)$ in the expressions for the interaction strengths by $\ {\cal F'}(0)$. However, when the mean global activity $\bar S$ is of order one, we have to include the contribution from ${\cal F}_0$ as well as modulations of interaction strengths due to $ \ {\cal F'}(n\bar S)$. In this case we can rewrite the transition rates as
	\begin{eqnarray}
		\omega (0 \to 1) &=& [\alpha_1 +  \beta'_1 {\cal F}_0] + \beta'_1  \  {\cal F}(\sum_{i\pm} S)  \nonumber\\
		&=& \alpha^{\text{eff}}_1 + \beta_1  \left(   \ \sum_{i\pm} S  \right)   \ ,
		\label{omega_01}
	\end{eqnarray}
	\begin{eqnarray}
		\omega (1 \to 0)  &=& [\alpha_2-  \beta'_2  {\cal F}_0 ]  - \beta'_2   \ {\cal F}(\sum_{i\pm} S)   \nonumber\\
		&=& \alpha^{\text{eff}}_2 - \beta_2  \left(   \ \sum_{i\pm} S  \right)   \ .
	   \label{omega_10}
	\end{eqnarray}
Thus, the effective intrinsic transition rates are activity dependent:
\begin{equation}
 \alpha^{\text{eff}}_1  =\alpha_1 +  \beta'_1 {\cal F}_0 \ , 
\end{equation}
	\begin{equation}
	\alpha^{\text{eff}}_2  =\alpha_2 -  \beta'_2 {\cal F}_0 \ , 
	\end{equation}
	
With these effective intrinsic transition rates, the intrinsic timescale also becomes activity dependent. Based on Eq.~\ref{equ:t0}, the equation for the intrinsic timescale can be rewritten as
	\begin{equation}
		\tau'_0 =\frac{1}{\alpha^{\text{eff}}_1+ \alpha^{\text{eff}}_2}= \frac{1}{\alpha_1+\alpha_2 +(\beta'_1-\beta'_2){\cal F}_0} \ .
		\label{equ:tau_0_act}
	\end{equation}
According to this equation, increasing mean global activity $\bar S$ leads to a decrease of  $\tau'_0$ when $(\beta'_1-\beta'_2)>0$ and to an increase of  $\tau'_0$ when $(\beta'_1-\beta'_2)<0$.
The changes of the intrinsic timescale result from a non-linear activation function and large values of the mean global activity $\bar S$. In the linear networks, ${\cal F}_0$ is zero and the intrinsic timescale is constant.

\subsection{Influence of external input on the operating regime of network dynamics and interaction timescales}

In models with non-linear interactions, the linear response of the system (derivative of the activation function) depends on the mean global activity and inputs. 
We show that for fixed interaction strength, the external input current changes the operating regime of network dynamics, which affects the magnitude of the mean global activity, its stability, and the intrinsic and interaction timescales.

The activation function with the external input is defined as:
	\begin{equation}
	 {\cal F} (n\bar S+I)  = 1- \exp(-\theta  \bar S-I) \ ,
	\end{equation}
where  $I$ represents a constant global input current (here we only consider the case $I \geqslant 0$).
In the steady-state, the mean global activity $\bar S$ follows the equation:
	\begin{eqnarray}
		\bar S   &\approx& \frac{\langle \omega(0\to1) \rangle}{\langle\omega(0\to1) \rangle+ \langle\omega(1\to0)\rangle}  \nonumber\\
		&\approx& \frac{\alpha_1 + \beta'_1 {\cal F}(n\bar S+I) }{\alpha_1 + \alpha_2 +  (\beta'_1 -\beta'_2) {\cal F}(n\bar S + I ) } \  .
		\label{sol}
	\end{eqnarray}
Since $0<{\cal F} \leqslant 1 $ and ${\cal F}(\infty) = 1$, in the large input limit $I \to \infty$, we have $\bar S (I \to \infty )= \frac{\alpha_1 + \beta'_1 }{\alpha_1 + \alpha_2 + (\beta'_1-\beta'_2)}$.

\begin{figure}[b]
\includegraphics{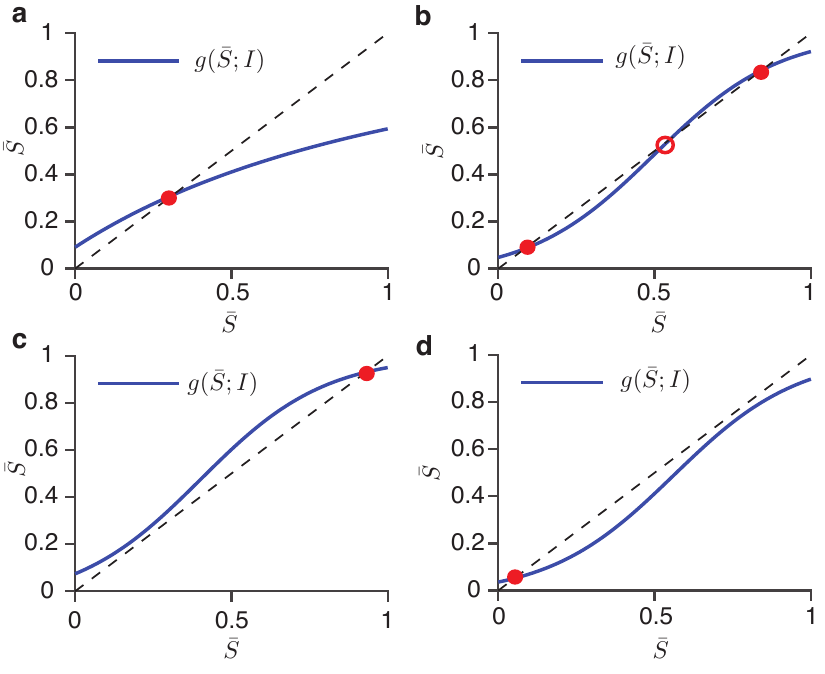}
\caption{\label{fig:phase}  Possible solutions for the global network activity.
In the $(x,y)$ plane, the intersections between the curve $x= \bar S, y=g(\bar S; I )$ (blue line) and the straight line $x=\bar S, y=\bar S$ (dashed line) are solutions of Eq.~\ref{sol} for the mean global activity $\bar S$.
(a) Only one stable solution (red dot). $\alpha_1=0.05/\Delta t$, $\alpha_2=0.5/\Delta t$, $\beta_1=1/\Delta t$, $\beta_2=0.5/\Delta t$, $I=0$, $\theta=0.5$. 
(b) Two stable solutions (filled dots) and one unstable solution (empty dot). $\alpha_1=0.025/\Delta t$, $\alpha_2=0.5/\Delta t$, $\beta_1=0.015/\Delta t$, $\beta_2=0.5/\Delta t$,  $I=0$, $\theta=0.5$.
(c) One stable solution (red dot)  in the sub-linear region.
$\alpha_1=0.04/\Delta t$, $\alpha_2=0.5/\Delta t$, $\beta_1=0.024/\Delta t$, $\beta_2=0.5/\Delta t$,  $I=0$, $\theta=5$.
(d) One stable solution (red dot) in the super-linear region.
$\alpha_1=0.017/\Delta t$, $\alpha_2=0.45/\Delta t$, $\beta_1=0.01/\Delta t$, $\beta_2=0.45/\Delta t$,  $I=0$, $\theta=5$.
}
\end{figure}

Eq.~\ref{sol} can have different solutions for $\bar S$ depending on the sign of $(\beta'_1-\beta'_2)$. To find these solutions, we define the function $g(\bar S;I)$:	
	\begin{equation}
		g(\bar S;I) =  \frac{\alpha_1 + \beta'_1 {\cal F}(n\bar S + I ) }{\alpha_1 + \alpha_2 +  (\beta'_1 -\beta'_2) {\cal F}(n\bar S + I ) } \  .
	\end{equation}
The solutions of Eq.~\ref{sol} are the intersections between the curve $x= \bar S, y=g(\bar S; I )$ and the straight line $x=\bar S, y=\bar S$ in the $(x,y)$ plane (Fig.~\ref{fig:phase}). The number and locations of the intersections depend on the first and second derivatives of $g(\bar S;I)$. The first derivative of $g(\bar S;I)$ is
	\begin{equation}
		g'(\bar S;I) =  \frac{[\beta'_1(\alpha_1+\alpha_2)-\alpha_1(\beta'_1-\beta'_2)]    {\cal F'}(n\bar S + I ) }{[\alpha_1 + \alpha_2 +  (\beta'_1 -\beta'_2) {\cal F}(n\bar S + I )]^2} \  ,
	\end{equation}
and the second derivative is 
\begin{widetext}
	\begin{eqnarray}
		g''(\bar S;I) &=&  [\beta'_1(\alpha_1+\alpha_2)-\alpha_1(\beta'_1-\beta'_2)]   \frac{   {\cal F'}(n\bar S+ I )[ -2(\beta'_1-\beta'_2)   {\cal F'}(n\bar S + I) -\theta (\alpha_1+\alpha_2+(\beta'_1-\beta'_2)    {\cal F}(n\bar S+I )   )  ] }{[\alpha_1 + \alpha_2 +  (\beta'_1 -\beta'_2) {\cal F}(n\bar S + I )]^3}  \nonumber\\
		&=&  [\beta'_1(\alpha_1+\alpha_2)-\alpha_1(\beta'_1-\beta'_2)]   \frac{   {\cal F'}(n\bar S+ I )[ -(\beta'_1-\beta'_2)   {\cal F'}(n\bar S+ I ) -\theta (\alpha_1+\alpha_2+(\beta'_1-\beta'_2)     )  ] }{[\alpha_1 + \alpha_2 +  (\beta'_1 -\beta'_2) {\cal F}(n\bar S +I )]^3}  \ .
	\end{eqnarray}
\end{widetext}
To determine the stability of solutions for $\bar S$, we consider a small deviation $\delta \bar S $ around the solution $\bar S=g(\bar S; I)$. The magnitude of fluctuation of the mean global activity is equal to $g'(n\bar S+I)\delta \bar S $.

Using the solution for $\bar S$, we can determine the effect of external input on the intrinsic (Eq.~\ref{equ:tau_0_act}) and interaction timescales. For simplicity, we consider only a representative interaction timescale, the global timescale, which is the largest interaction timescales. For a given mean global activity $\bar S $ and external input $I$, the global timescale  is
\begin{eqnarray}
&& \tau_{\text{global}} = \frac{\tau'_0}{1- \frac{n\beta_1}{\alpha^{\text{eff}}_1+ \alpha^{\text{eff}}_2}} \nonumber \\ 
&&=   \frac{\tau'_0 }{1-  \frac{n \beta'_1   {\cal F'}(n\bar S+I)}{\alpha_1+\alpha_2+ (\beta'_1-\beta'_2){\cal F}_{0} }}   \nonumber\\
	&&= \frac{1}{\alpha_1+\alpha_2+(\beta'_1-\beta'_2){\cal F}_0-n\beta'_1 {\cal F'}(n\bar S+I) } \nonumber \\
	&&=\frac{1}{\alpha_1+\alpha_2+(\beta'_1-\beta'_2)[1-(\theta \bar S +1)e^{-\theta \bar S-I}]-\beta'_1 \theta e^{-\theta \bar S-I} } \ . \nonumber\\
	\end{eqnarray}
To study the dependence of $\tau_{\text{global}}$ on $\bar S$, we take the derivative of the denominator with respect to $\bar S$:
	\begin{equation}
	    \frac{d}{d\bar S} \tau_{\text{global}} \propto
	   -[ (\beta'_1-\beta'_2) \bar S +\beta'_1 ] \theta^2 e^{-\theta \bar S-I } \ .
	\end{equation}
Since both timescales $\tau'_0$ and $\tau_{\text{global}}$ depend on $\bar S$, the external input can affect the timescales by changing $\bar S$. If ${\cal F}$ is a linear function of $\bar S$, then both ${\cal F}_0$ and  ${\cal F}'$ are independent of $\bar S$ and hence the input does not influence the timescales \cite{Huang2017}.

Depending on the parameters $\alpha_{1,2}, \beta_{1,2},\theta$, there are two classes of solutions for $\bar S$ depending on the sign of $(\beta'_1 -\beta'_2)$. In the following, we discuss these possible solutions and how they affect the intrinsic and global timescales.

		\subsubsection{$(\beta'_1 -\beta'_2)>0$}
	
When $(\beta'_1 -\beta'_2)>0$, since parameters  $\alpha_{1,2}, \beta_{1,2}$ and $\theta$ are positive definite, ${	g''(\bar S;I ) }$ and ${	g'(\bar S;I) } $ have opposite signs:
	\begin{eqnarray}
&& \frac{	g''(\bar S;I) }{	g'(\bar S;I) }  \nonumber \\
&=& \frac{[ -(\beta'_1-\beta'_2)   {\cal F'}(n\bar S+I ) -\theta (\alpha_1+\alpha_2+(\beta'_1-\beta'_2)     )  ] }{[\alpha_1 + \alpha_2 +  (\beta'_1 -\beta'_2) {\cal F}(n\bar S+I)]^3}  <0 \ , \nonumber\\
	\end{eqnarray}
which means that $g(\bar S;I )$ always has a sub-linear behavior. The absolute value of the derivative $\bar g'(\bar S;I)$ exponentially decreases with increasing $n\bar S+I$ and approaches zero in the asymptotic limit.
In this case, the first derivative is positive $\bar g'(\bar S;I)>0$, and the second derivative is negative $\bar g''(\bar S;I)<0$, hence the asymptotic value of $\bar S (I \to \infty )\equiv g (\bar S=1;I)$ is larger than the non-interaction component $\alpha_1/(\alpha_1+\alpha_2)$ (which is $g(S=0;I)$).
		
In this configuration, there is only one solution for $\bar S$, in the range from $\alpha_1/(\alpha_1+\alpha_2)$ to $\bar S (I \to \infty )$ (Fig. \ref{fig:phase}a). With increasing current $I$, the global activity $\bar S$ increases, leading to a reduction in both the intrinsic timescale $\tau'_0$ and the global interaction timescale $\tau_{\text{global}}$.

	\subsubsection{$(\beta'_1 -\beta'_2)<0$}
	
When $(\beta'_1 -\beta'_2)<0$, the first derivative is positive $ g'(\bar S;I) >0 $, so $g(\bar S;I)$ is an increasing function of $n\bar S+I$. Depending on the interaction strength $\theta$, $ g''(\bar S;I) $ can be positive or negative. Hence, we classify the operating regime of activity $\bar S$ based on the sign of $ g''(\bar S;I) $:

	\begin{enumerate}[label=\alph*)]
		\item  Two stable solutions of $\bar S$ and one unstable solution (Fig.~\ref{fig:phase}b).
	   If $	g'(0;I) > 1 $,  and there are two solutions for the equation 
		\begin{equation}
			g'(\bar S;I) -1 =0  |_{\bar S= \bar S_1,  \bar S_2}   \ ,
		\end{equation}
		where $\bar S_1$ and $\bar S_2$ are two solutions of the equation $g(\bar S) =\bar S$, and they satisfy the constraints:
		\begin{equation}
		g(\bar S_1, I)  < \bar S_1 -I \  ; \quad   	g(\bar S_2, I)  < \bar S_2 -I  \ ,
		\end{equation}
		then, there are three solutions for $\bar S $. One stable solution within $[0, S_1]$, where  $ g''(\bar S;I)  >0$ (supra-linear),  one unstable solution within $[S_1, S_2]$ and one stable solution within $[S_2, 1]$, where  $ g''(\bar S;I)  <0$ (sub-linear).

		\item One solution for $\bar S$ in the sub-linear or supra-linear region (Fig.~\ref{fig:phase}c,d).
		When conditions in case (a) are not satisfied, there is always one solution for $\bar S$ in sub-linear or supra-linear region.
		
	\end{enumerate}	
	
Considering these different solutions, with increasing $I$, the mean global activity $\bar S$ increases, leading to an increase in  intrinsic timescale $\tau'_0$. When $-|\beta'_1-\beta'_2| \bar S +\beta'_1>0$,  $\tau_{\text{global}}$ decreases with $I$. When $-|\beta'_1-\beta'_2| \bar S +\beta'_1<0$, $\tau_{\text{global}}$ increases with $I$.

\section{Discussion}

We studied the spatial and temporal scales of neural correlations in binary-unit networks with connectivity arranged in one- and two-dimensional space. We used the time-evolution equations for correlation functions derived from the master equation. We solved these equations using the discrete Fourier transform and translational symmetry of the model and obtained analytical solutions for spatiotemporal correlations. We  found that the spatial and temporal scales of correlations are related to each other and shaped by the spatial profile of the recurrent connectivity. Finally, we showed that external inputs can control the operating regime of the global network activity and thus influence the timescales of correlations. To confirm our theoretical results, we performed numerical simulations and found a good agreement between analytical solutions and simulation results.

One of our key findings is that spatial recurrent interactions generate multiple timescales in network dynamics. The spatial interactions we considered are similar to the spatial connectivity structure in the primate cortex. The distance-dependent connectivity perseveres the translational symmetry, hence in Fourier space, each spatial Fourier mode of correlations is approximately decoupled and evolves with a unique characteristic interaction timescale. In the strong interaction limit, the interaction timescales can be significantly larger than the intrinsic timescale. The overall temporal profile of correlations arises from a superposition of all Fourier modes with distinct timescales. These interaction timescales depend on the spatial range of connectivity in heterogeneous manner. In particular, local spatial connectivity tends to enhance a broad spectrum of interaction timescales, while homogeneous all-to-all connectivity eliminates all interaction timescales except for the global timescale associated with the spatially homogeneous component of correlations. Therefore, in our network models, multiple timescales are inherently coupled to the spatial connectivity, which is different from other models where heterogeneous timescales are generated by single-cell proprieties such as  self-couplings \cite{Stern2022}. The relation between timescales and connectivity has been analyzed in a deterministic linear network model \cite{Chaudhuri2014}, where timescales are defined by the eigenvalues of the connectivity matrix. Here we study the relation between structural connectivity and timescales of neural correlations in stochastic networks of binary-units.

Another major contribution of our work is to establish the link between spatial and temporal scales of correlations. Our theory predicts that slow interaction timescales in autocorrelations of networks with spatial connectivity are generated by correlations between the activity of units at different distances. In these networks, correlations at different distance have distinct amplitude spectra of their spatial Fourier modes. Since each Fourier mode carries a unique interaction timescale, the overall temporal structure of correlations depends on the the spatial distance. In particular, the average interaction timescale tends to be larger for correlations between pairs of neurons with a larger distance. This feature is supported by a recent analysis of spiking activity in primate visual cortex \cite{Zeraati2021}.

We showed that when the interaction between the network units is non-linear, the external input current changes the operating regime of network dynamics and modifies the intrinsic and interaction timescales of correlations. This mechanism of modulating timescales through external input may have implications in biological circuits. For example, in neocortex top-down inputs from higher cortical areas can regulate dynamics of cortical states in sensory areas \cite{Harris2011}, which may modulate timescales of fluctuations \cite{Engel2016, Shi2022, Zeraati2021}.

In this paper, we considered models with spatial connectivity patterns where each unit connects to all its neighbors within a radius $R$. In the future, it would be interesting to extend the current framework to study network models with random spatial connectivity. In this case, the connectivity patterns can be described by random band matrices. According to theories of random band matrices \cite{Randomband}, the spatial correlation undergoes a transition from localization to delocalization phases, when the range of spatial connectivity exceeds certain thresholds. In addition, we focused here on the regime with a stable activity. Another future extension is to explore the spatiotemporal correlations in the dynamical regime where rate dynamics are chaotic \cite{Sompolinsky1988,Landau2018}.

\begin{acknowledgments}
This work was supported by the Swartz Foundation (Y.S.), a Sofja Kovalevskaja Award from the Alexander von Humboldt Foundation, endowed by the Federal Ministry of Education and Research (R.Z., A.L.), SMARTSTART2 program provided by Bernstein Center for Computational Neuroscience and Volkswagen Foundation (R.Z.), the NIH grant RF1DA055666 (T.A.E.), the Feil Family Foundation (T.A.E.), and Alfred P. Sloan Foundation Research Fellowship (Y.S., T.A.E.).
\end{acknowledgments}

Y.S., R.Z., A.L., and T.A.E. designed the study.  Y.S., R.Z., A.L., and T.A.E. developed the model and analysis methods. Y.L. performed the analytical calculations. R.Z. performed model simulations. Y.S., R.Z., A.L., and T.A.E. wrote the paper.

\appendix

 	\section{Relation between the continuous-time analytical model and discrete-time simulations}
	\label{simulation}
	 
	 In simulations of binary-unit network models, we update the state of units based on transition probabilities at discrete time steps. In the analytical calculations, on the other hand, we describe the dynamics using the instantaneous transition rates in continuous time. Here, we discuss how these two different representations of dynamics are related to each other.
	 
	 In discrete-time dynamics, the state of a binary unit $S_i \in \{0,1\}$ is updated at each time step $\Delta t$ based on the transition probabilities, which depend on the sum of states of its directly connected neighbours (denoted by $\sum_{j} S_j$):
	\begin{equation}
		p (0 \to 1; S_i=0) = p_{\text{ext}}+  	{\cal F}(\sum_{j} S_j) \   , \label{Eq:A1}
	\end{equation}
	\begin{equation}
	p (0 \to 0; S_i=0) = 1 -	p (0 \to 1; S_i=0) \ .\label{Eq:A2}
	\end{equation}
	\begin{equation}
		p (1 \to 0; S_i=1) = 1- p_{\text{ext}}-p_s- 	{\cal F}(\sum_{j} S_j)   \ ,
	\end{equation}
	\begin{equation}
	p (1 \to 1; S_i=1) = 1 -	p (1 \to 0; S_i=1) \ .\label{Eq:A4}
	\end{equation}
	Here, we define the interaction term as a linear function 
	\begin{equation}
		{\cal F}(\sum_{j} S_j) =  p_r \cdot \left( \sum_{j} S_j \right)  \ .\label{Eq:A5}
	\end{equation}

    Generally, ${\cal F}$ should satisfy the condition  ${\cal F}(0) =0 $, ${\cal F}(\infty) =1  $, and ${\cal F}(x) $ is a monotonically increasing function of $x$. When the mean global activity of the network is much smaller than 1, the above linear definition serves as a good approximation. Thus, for each unit we define a transition matrix between binary states:
		\begin{equation*}
	    \begin{pmatrix}
	     P(S_i(t+\Delta t) = 0| S_i(t) = 0) &    P(S_i(t+\Delta t) = 1|S_i(t) = 0) \\
	      P(S_i(t+\Delta t) = 0| S_i(t) = 1) &   P(S_i(t+\Delta t) = 1| S_i(t) = 1)  \\ 
	    \end{pmatrix}.
	\end{equation*}
Using equations~\ref{Eq:A1}--\ref{Eq:A4}, we can write the transition matrix as:  
	\begin{equation}
	P(\Delta t) =
	    \begin{pmatrix}
	      1-p_{\text{ext}}   -	{\cal F} &  p_{\text{ext}}+	{\cal F} \\
	     1-p_{\text{ext}}-p_s-	{\cal F}  &  p_s+  p_{\text{ext}} +  {\cal F}  \\ 
	    \end{pmatrix} \ .
	\end{equation}

In our analytical calculations of binary-unit dynamics, we use the instantaneous transition rates $\alpha_{1,2}$ and $\beta_{1,2}$ to describe the changes in the probability density of the states. To link transition rate parameters to transition probabilities ($p_{\text{ext}}$, $p_{s}$, $p_{r}$), we use the fact that the transition matrix $P(\Delta t)$ can be approximated by the matrix exponential of transition rate matrix $e^{Q\Delta t}$, where the transition rate matrix $Q$ is given by 
\begin{equation}
    Q= \begin{pmatrix}
     -	\omega (0 \to 1) & 	\omega (0 \to 1) \\
      	\omega (1 \to 0) & 	- \omega (1 \to 0) \\ 
    \end{pmatrix} \ .
\end{equation}
Here, $\omega (0 \to 1) $ describes the transition rate from state $0$ to $1$, and  $\omega (1 \to 0)$  describes the transition rate from state $1$ to $0$. Then, the matrix exponential of transition rate matrix can be written as
	\begin{widetext}
	\begin{equation}
	e^{Q\Delta t} =
	    \begin{pmatrix}
	     \frac{\omega (1 \to 0) }{\omega (0 \to 1) +\omega (1 \to 0) } + \frac{\omega (0 \to 1) }{\omega (0 \to 1) +\omega (1 \to 0) } e^{-[\omega (0 \to 1) +\omega (1 \to 0) ]\Delta t} &  	     \frac{\omega (0 \to 1) }{\omega (0 \to 1) +\omega (1 \to 0) } - \frac{\omega (0 \to 1) }{\omega (0 \to 1) +\omega (1 \to 0) } e^{-[\omega (0 \to 1) +\omega (1 \to 0) ]\Delta t}\\
	   	     \frac{\omega (1 \to 0) }{\omega (0 \to 1) +\omega (1 \to 0) } - \frac{\omega (1 \to 0) }{\omega (0 \to 1) +\omega (1 \to 0) } e^{-[\omega (0 \to 1) +\omega (1 \to 0) ]\Delta t}  &  	     \frac{\omega (0 \to 1) }{\omega (0 \to 1) +\omega (1 \to 0) } + \frac{\omega (1 \to 0) }{\omega (0 \to 1) +\omega (1 \to 0) } e^{-[\omega (0 \to 1) +\omega (1 \to 0)]\Delta t}  \\ 
	    \end{pmatrix} \ .
	\end{equation}
	\end{widetext}
Solving $e^{Q\Delta t}= P(\Delta t)$ and using Eq.~\ref{Eq:A1}--\ref{Eq:A5}, we have 
	\begin{eqnarray}
		\omega (0 \to 1) &=&   [ p_{\text{ext}}  + {\cal F}(\sum_{j} S_j) ] \left [ \frac{-\ln p_s}{(1-p_s)\Delta t} \right ] \nonumber \\
		&=& \alpha_1  + \beta'_1 \ {\cal F}(\sum_{j} S_j)  \nonumber \\
		&=&  \alpha_1  + \beta_1 \cdot \left(\sum_{j} S_j \right) \ ,
	\end{eqnarray}
	where, transition rates $\alpha_1, \ \beta_1$ are given by: 
	\begin{equation}
		\alpha_1 =  p_{\text{ext}}  \left [ \frac{-\ln p_s}{(1-p_s)\Delta t} \right ]   \ ,
	\end{equation}
		\begin{equation}
	    \beta_1 =  p_r \left [ \frac{-\ln p_s}{(1-p_s)\Delta t} \right ]  \ ,
	\end{equation}
	and 
	\begin{eqnarray}
	\omega (1 \to 0)   &=&   [(1-p_s-p_{\text{ext}}  - {\cal F}(\sum_{j} S_j) ]  \left [ \frac{-\ln p_s}{(1-p_s)\Delta t} \right ]  \nonumber \\
		&=& \alpha_2  - \beta'_2   \ {\cal F}(\sum_{j} S_j) =  \alpha_2  - \beta_2  \cdot \left(\sum_{j} S_j\right)  \ , \nonumber\\
	\end{eqnarray}
	where transition rates $\alpha_2, \ \beta_2$ are given by 
	\begin{equation}
		\alpha_2 =    (1-p_s-p_{\text{ext}})  \left [ \frac{-\ln p_s}{(1-p_s)\Delta t} \right ]    \ .
	\end{equation}
		\begin{equation}
	   \beta_2 =  p_r   \left [ \frac{-\ln p_s}{(1-p_s)\Delta t} \right ]   \ .
	\end{equation}
	We see that for the transition rates corresponding to the parameters of discrete model, $\beta_1 = \beta_2$.

	\section{Derivation of dynamical equations for the moments}
	\label{appendix3}
	
	We denote the probability of the network to be in a certain configuration  $\{S\}=\{S_1, S_2,...,S_N\}$ at time $t$ by  $P(\{S\},t)$. The master equation describes the time evolution of $P(\{S\},t)$, which is given by \cite{Ginzburg1994,Renart2010,Chow2010,Dahmen2016,Farkhooi}:
	\begin{eqnarray}
		&&\frac{d}{dt}P(\{S\},t)  \nonumber\\
		&=& - P(\{S\},t) \sum_{i} w(S_i)+ \sum_{i}  P(\{S\}^{i*},t) w(1- S_i) \ , \nonumber\\
	\end{eqnarray}
	where $\{S\}^{i*}=\{S_1, S_2,...,1-S_i,...,S_N\}$ and $w(S_i)$ is the transition rate from state $S_i$ to $1-S_i$.

	Using the master equation, one can write down the equation for the time evolution of arbitrary moments. For example, the average activity of a unit $i$ is defined as
	\begin{equation}
		\langle S_i \rangle(t) = \sum_{\{S\}}P(\{S \},t) S_i \ ,
		\label{average_triangle}
	\end{equation}
	where we sum over all configurations of variables $\{S\}$ at a given time $t$. The time evolution of the average activity is given by
	\begin{eqnarray}
		\frac{d}{dt}\langle S_i \rangle(t)&=& \frac{d}{dt} \left(\sum_{\{S\}}P(\{S \},t) S_i \right) \nonumber\\
		&=& \sum_{\{S\}} \left(\frac{d}{dt} P(\{S \},t) \right) S_i \ .
	\end{eqnarray}
	Substituting the master equation, we have 
	\begin{equation}
		\frac{d}{dt}\langle S_i \rangle(t)= \sum_{\{S\}} P(\{S\},t) [w(S_i)(1-2S_i)] \ .
	\end{equation}
	Similarly, the rate of change of a second moment for each pair of units is
	\begin{eqnarray}
		&&\frac{d}{dt}\langle S_i S_j \rangle(t) \nonumber\\
		&=& \sum_{\{S\}} P(\{S\},t) [w(S_i)(1-2S_i)S_j+ w(S_j)(1-2S_j)S_i] \ . \nonumber \\
	\end{eqnarray}
	
	The time evolution of time-delayed second moment can be computed as 		\cite{Ginzburg1994,Renart2010,Chow2010,Dahmen2016,Farkhooi}:
	\begin{eqnarray}
	 &&	\frac{d}{d\tau}\langle S_i (t) S_j (t+\tau )\rangle  \nonumber\\
		&=&  \sum_{\{S\}}P(\{S \},t) S_i \frac{d}{d\tau} \left(\sum_{\{\sigma\}}P(\{\sigma \},t+\tau|\{S \},t) \sigma_j \right) \ . \nonumber\\
	\end{eqnarray}
	where $P(\{\sigma \},t+\tau|\{S \},t)$ is conditional probability of finding the system in configuration $\{\sigma\}$ at time $t+\tau$, given that it
	was in configuration $\{S\}$ at time $t$. Since the conditional probability obeys the same master equation, we have
	\begin{equation}
		\frac{d}{d\tau}\langle S_i (t) S_j (t+\tau )\rangle = \langle S_i(t) (1-2S_j(t+\tau)) w(S_j(t+\tau))  \rangle  \ .
	\end{equation}

	Substituting the explicit form of the transition rates and summing over all configurations, we get the following coupled equations for the first moment \cite{Ginzburg1994,Renart2010,Chow2010,Dahmen2016,Farkhooi}:
	\begin{eqnarray}
	 \frac{d}{dt}\langle S_i \rangle(t) &=& \alpha_1-(\alpha_1 + \alpha_2) \langle S_i \rangle \nonumber\\
		&+& \beta_1 \langle \sum_{l; \, l \to i} S_l \rangle  + (\beta_2-\beta_1) \langle S_i \sum_{l;\, l \to i} S_l \rangle  \ . 
	\end{eqnarray}
	Here, $\sum_{l; \, l \to i} S_l$ denotes the sum of states of units directly connected to unit $i$. Subtracting the mean $\delta S_i = S_i - \langle S_i \rangle$, we find the time-evolution equation for equal-time correlation as

		\begin{eqnarray}
		&& \frac{d}{dt}\langle \delta S_i (t) \delta S_j (t)\rangle = - 2(\alpha_1+\alpha_2)\langle \delta S_i \delta  S_j \rangle \nonumber\\
		&+& \beta_1 (\langle  \sum_{l; \, l \to i} \delta S_l \cdot \delta S_j\rangle + \langle \delta S_i \sum_{l; \, l \to j} \delta S_l \rangle) \nonumber\\
		&+& (\beta_2-\beta_1) (\langle \delta S_i \sum_{l; \, l \to i} \delta S_l \delta S_j\rangle + \langle  \delta S_i \delta S_j \sum_{l; \, l \to j} \delta S_l    \rangle ) \ , \  (i \neq j) \nonumber\\ 
	\end{eqnarray}

	Substituting the explicit form of transition rates into the time-evolution of time-delayed quadratic moment, we find the time-evolution equation for autocorrelation
	\begin{eqnarray}
		&&\frac{d}{d\tau}\langle \delta S_i (t) \delta S_i (t+\tau)\rangle \nonumber\\
		&=& - (\alpha_1+\alpha_2)\langle  \delta S_i (t) \delta S_i (t+\tau )\rangle  \nonumber\\
		&+& \beta_1   \langle \delta S_i (t) \sum_{l; \, l \to i}  \delta S_l(t+\tau)  \rangle 
		\nonumber\\
		&+& (\beta_2-\beta_1) (\langle \delta S_i  (t)    \delta S_i (t+\tau)    \sum_{l; \, l \to i} \delta S_l (t+\tau)  \rangle  ) \  . \nonumber\\
	\end{eqnarray}
	and the time-evolution equation for the time-delayed cross-correlation 
	\begin{eqnarray}
		&& \frac{d}{d\tau}\langle \delta S_i (t) \delta S_j (t+\tau) \rangle  =  \nonumber\\
		& -& (\alpha_1+\alpha_2)\langle  \delta S_i (t) \delta S_j (t+\tau )\rangle + \beta_1   \langle  \delta S_i (t)  \sum_{l; \, l \to j}  \delta S_l(t+\tau)  \rangle 
		\nonumber\\
		&+& (\beta_2-\beta_1) (\langle \delta S_i  (t)    \delta S_j (t+\tau)    \sum_{l; \, l \to j}  \delta S_l(t+\tau)  \rangle  ) \ , \  (i \neq j) \ . \nonumber\\
	\end{eqnarray}

   \section{Time evolution of averaged correlation functions in two-dimensional model}
	\label{2d_equatoin}
	
For the two-dimensional models with nearest-neighbor connectivity, the steady state equation for equal-time cross-correlation function is given by
\begin{eqnarray}
 &&C_2(x_1,x_2) = \frac{\beta_1}{\alpha_1+\alpha_2 }[ C_2(x_1-a,x_2)  \nonumber \\
 &&+ C_2(x_1+a,x_2) + C_2(x_1,x_2+a) + C_2(x_1,x_2-a) \nonumber \\
 &&+ C_2(x_1+a,x_2+a) + C_2(x_1+a,x_2-a)   \nonumber \\
 &&+ C_2(x_1-a,x_2+a) + C_2(x_1-a,x_2-a)  \nonumber \\
 && + ( \delta_{x_1,0} \delta_{x_2,a}  + \delta_{x_1,0} \delta_{x_2,-a}  + \delta_{x_1,-a} \delta_{x_2,0} \nonumber\\
 && +  \delta_{x_1,a} \delta_{x_2,0}  + \delta_{x_1,a} \delta_{x_2,a}  + \delta_{x_1,a} \delta_{x_2,-a} \nonumber\\
 &&+ \delta_{x_1,-a} \delta_{x_2,a} + \delta_{x_1,-a} \delta_{x_2,-a}) A (0)  ] \ .
 \label{cc_r1_2d}
\end{eqnarray}	
The time-evolution equation for the time-delayed cross-correlation function is 	
\begin{eqnarray}
&&\tau_0 \frac{d}{dt} C_2(x_1,x_2,t) =  - C_2(x_1,x_2,t)  \nonumber \\
&&+ \frac{\beta_1}{\alpha_1+\alpha_2 } [ C_2(x_1-a,x_2,t)  + C_2(x_1,x_2-a,t)   \nonumber \\
 &&+ C_2(x_1+a,x_2,t) + C_2(x_1,x_2+a,t) \nonumber \\
 &&+ C_2(x_1+a,x_2+a,t) + C_2(x_1+a,x_2-a,t)   \nonumber \\
 &&+ C_2(x_1-a,x_2+a,t) + C_2(x_1-a,x_2-a,t)  \nonumber \\
 && + ( \delta_{x_1,0} \delta_{x_2,a}  + \delta_{x_1,0} \delta_{x_2,-a}  + \delta_{x_1,-a} \delta_{x_2,0} \nonumber\\
 && +  \delta_{x_1,a} \delta_{x_2,0}  + \delta_{x_1,a} \delta_{x_2,a}  + \delta_{x_1,a} \delta_{x_2,-a} \nonumber\\
 &&+ \delta_{x_1,-a} \delta_{x_2,a} + \delta_{x_1,-a} \delta_{x_2,-a}) A_2 (t)  ] \ .
 \label{cc_t_r1_2d}
\end{eqnarray}		
The time-evolution equation for the average autocorrelation function is 	
\begin{eqnarray}
&&\tau_0 \frac{d}{dt} A_2(t) =  - A_2(t) +\frac{\beta_1}{\alpha_1+\alpha_2 } [ 4 C_2(a,a,t)  \nonumber \\
&& + 2 C_2(a,0,t) + 2 C_2(0,a,t)].
 \label{ac_t_r1_2d}
\end{eqnarray}	
In Fourier space, the steady state equation for $\tilde  C_2(k_1,k_2)  $ is given by 
\begin{eqnarray}
&& \tilde  C_2(k_1,k_2)     =  \nonumber \\
&&\frac{2 \beta_1}{\alpha_1+\alpha_2} [ \cos(k_1 a) + \cos(k_2 a) + 2\cos(k_1 a)  \cos(k_2 a)] \nonumber\\
&& \times \tilde C_2(k_1,k_2) + \frac{2 \beta_1}{\alpha_1+\alpha_2}  \frac{4}{N^2} \nonumber \\
&&\times [ \cos(k_1 a) + \cos(k_2 a) + 2\cos(k_1 a)  \cos(k_2 a)] A(0)   .
\end{eqnarray}
The time-evolution equation for $\tilde  C_2(k_1,k_2,t)  $ is 
\begin{eqnarray}
&&\tau_0 \frac{d}{dt} \tilde C_2(k_1,k_2,t) = -    C_2(k_1,k_2,t)     \nonumber \\
&&+\frac{2 \beta_1}{\alpha_1+\alpha_2} [ \cos(k_1 a) + \cos(k_2 a) + 2\cos(k_1 a)  \cos(k_2 a)] \nonumber\\
&& \times \tilde C_2(k_1,k_2,t) + \frac{2 \beta_1}{\alpha_1+\alpha_2}  \frac{4}{N^2}\nonumber \\
&&\times [ \cos(k_1 a) + \cos(k_2 a) + 2\cos(k_1 a)  \cos(k_2 a)] A(t) \nonumber \\
&& \approx -\frac{\tau_0}{\tau(k_1,k_2)} \tilde C_2(k_1,k_2,t) \ .
\end{eqnarray}


For the two-dimensional models with long-range connectivity ($R>1$), the steady state equation for equal-time cross-correlation function is given by
\begin{eqnarray}
 &&C_2(x_1,x_2;R) = \frac{\beta_1}{\alpha_1+\alpha_2 } \nonumber \\
&& \times \left[ \sum^{R}_{m_1,m_2=-R} C_2(x_1+m_1a,x_2+m_2a;R) -C_2(x_1,x_2;R) \right]  \nonumber \\
&& + \frac{\beta_1}{\alpha_1+\alpha_2 } \sum^{R}_{m_1,m_2=-R} [\delta_{x_1+m_1a,x_2+m_2a} ] A(0) \ .
 \label{cc_rr_2d}
\end{eqnarray}	
The time-evolution equation for the time-delayed cross-correlation function is 	
\begin{eqnarray}
 &&\tau_0 \frac{d}{dt} C_2(x_1,x_2,t;R) =  -C_2(x_1,x_2,t;R) + \frac{\beta_1}{\alpha_1+\alpha_2 } \nonumber \\
&& \times [ \sum^{R}_{m_1,m_2=-R} C_2(x_1+m_1a,x_2+m_2a,t;R)  \nonumber\\
&&\left.-C_2(x_1,x_2,t;R) \right]  \nonumber \\
&& + \frac{\beta_1}{\alpha_1+\alpha_2 } \sum^{R}_{m_1,m_2=-R} [\delta_{x_1+m_1a,x_2+m_2a} ] A(0) \ .
 \label{cc_t_rr_2d}
\end{eqnarray}		
The time-evolution equation for the autocorrelation function $A_2(t;R)$ is 	
\begin{eqnarray}
&&\tau_0 \frac{d}{dt} A_2(t;R) =  - A_2(t;R) +\frac{\beta_1}{\alpha_1+\alpha_2 }  \nonumber\\
&&\times [ \sum^{R}_{m_1,m_2=-R} C_2(m_1a,m_2a,t;R) ] \ .
 \label{ac_t_rr_2d}
\end{eqnarray}	
Solving the equations for $C_2(x_1,x_2)$, $C_2(x_1,x_2,t)$, $C_2(x_1,x_2;R)$, $C_2(x_1,x_2,t;R)$, and take the value $x_1=\Delta_1$, $x_2=\Delta_2$, we can get average correlations with fixed distance $(\Delta_1,\Delta_2)$: $C_2(\Delta_1,\Delta_2)$, $C_2(\Delta_1,\Delta_2,t)$, $C_2(\Delta_1,\Delta_2;R)$, $C_2(\Delta_1,\Delta_2,t;R)$.


\begin{thebibliography}{99}

	      
	      
	      \bibitem{Cohen2011}
	        M. R. Cohen and A. Kohn,  Measuring and interpreting neuronal correlations. Nat. Neurosci. 14, 811–819 (2011).
	        
	        
	       \bibitem{Smith2008}
          M. A. Smith and A. Kohn, Spatial and Temporal Scales of Neuronal Correlation in Primary Visual Cortex, Journal of Neuroscience, 28 (48) 12591-12603, (2008).
          
          
          \bibitem{Smith2013}
          M. A. Smith and M. A. Sommer, Spatial and Temporal Scales of Neuronal Correlation in Visual Area V4, Journal of Neuroscience,  33 (12) 5422-5432,  (2013).
          
          
          \bibitem{Huang2019}
	     C. Huang, D.A. Ruff, R. Pyle, R. Rosenbaum, M.R. Cohen, B. Doiron, Circuit models of low-dimensional shared variability in cortical networks, Neuron, 101 , 337-348 (2019).
          
          
          \bibitem{Shi2022}
	       Y.-L. Shi, N.A. Steinmetz, T. Moore, K. Boahen, T.A. Engel, Cortical state dynamics and selective attention define the spatial pattern of correlated variability in neocortex, Nature Communications 13 , 1-15 (2022).
          
           \bibitem{Dahmen2022}
           Dahmen, David et.al., Global organization of neuronal activity only requires unstructured local connectivity, eLife, 11, e68422  (2022).
          
	       \bibitem{Safavi2018}
           Safavi, S. et.al., Nonmonotonic spatial structure of interneuronal correlations in prefrontal microcircuits. Proceedings of the National Academy of Sciences, 115(15), E3539-E3548 (2018). 
           
           
           \bibitem{Murray2014}
	      J. D. Murray, et al., A hierarchy of intrinsic timescales across primate cortex, Nature Neuroscience 17, 1661–1663 (2014).
           
           
            \bibitem{Okun2019}
            M. Okun, N. A. Steinmetz, A. Lak, M. Dervinis, K. D. Harris, Distinct Structure of Cortical Population Activity on Fast and Infraslow Timescales, Cerebral Cortex 29, 2196 (2019)
           
            \bibitem{Gao2020}
	      R. Gao, R. L. van den Brink, T. Pfeffer, and B. Voytek, Neuronal timescales are functionally dynamic and shaped by cortical microarchitecture,  Elife 9, e61277 (2020).
	      
	      
	       \bibitem{Runyan2017}
            C. A. Runyan, E. Piasini, S. Panzeri and C. D. Harvey, Distinct timescales of population coding across cortex, Nature  548, 92–96 (2017).
            
            
	        

	      
	      
	      
	        \bibitem{Harris2011}
	        K. D. , Harris. and  A. Thiele, Cortical state and attention. Nat. Rev. Neurosci. 12, 509–523 (2011).
	        
	        
	        
	        \bibitem{Cohen2009}
	        M. R. Cohen and  J. H. R. Maunsell,  Attention improves performance primarily by reducing interneuronal correlations. Nat. Neurosci. 12, 1594–1600 (2009).

            \bibitem{Mitchell2009}
            J. F. Mitchell, K. A. Sundberg, J. H. Reynolds,  Spatial attention decorrelates intrinsic activity fluctuations in Macaque area V4. Neuron 63, 879–888 (2009).

            \bibitem{Ruff2016}
            D. A. Ruff,  M. R. Cohen, Attention increases spike count correlations between visual cortical areas. J. Neurosci. 36, 7523–7534 (2016).
 
            \bibitem{Nandy2017}
            A. S. Nandy, J. J. Nassi,  J. H. Reynolds, Laminar organization of attentional modulation in Macaque visual area V4. Neuron 93, 235–246 (2017).

             \bibitem{Denfield2018}
              G. H. Denfield, A. S. Ecker, T. J. Shinn, M. Bethge, and A. S. Tolias, Attentional fluctuations induce shared variability in macaque primary visual cortex. Nat. Commun. 9, 1–14 (2018).
	       
	       
	     \bibitem{Zeraati2021}
		 R. Zeraati,  Y.-L. Shi, N. A. Steinmetz,  M. A. Gieselmann, A. Thiele, T. Moore,  A. Levina,  T. A. Engel, Attentional modulation of intrinsic timescales in visual cortex and spatial networks, biorxiv (2021), doi: https://doi.org/10.1101/2021.05.17.444537
	      
	      
	      \bibitem{Bernacchia2011}
	        A. Bernacchia, H.g Seo, D. Lee and X.-J. Wang,  A reservoir of time constants for memory traces in cortical neurons, Nature Neuroscience  14, 66–372 (2011).     
            
            \bibitem{Cavanagh2016}
            S. E. Cavanagh, J. D.  Wallis, S. W. Kennerley and L. T. Hunt, Autocorrelation structure at rest predicts value correlates of single neurons during reward-guided choice, eLife 5, e18937 (2016).

            \bibitem{Wasmuht2018}
            D. F. Wasmuht, E. Spaak, T. J. Buschman, E. K. Miller, and M. G. Stokes, Intrinsic neuronal dynamics predict distinct functional roles during working memory, Nature Communications 9, 3499 (2018) 
        
            
        \bibitem{Mountcastle1997}
	      Mountcastle, V. B. The columnar organization of the neocortex. Brain 120, 701–722 (1997).
	      
	   \bibitem{Buxhoeveden2002}
       D. P. Buxhoeveden and M. F. Casanova,  The minicolumn hypothesis in neuroscience , Brain 125,  935–951 (2002).
            
          
	      \bibitem{Rosenbaum2017}
	     R. Pyle, R. Rosenbaum, Spatiotemporal dynamics and reliable computations in recurrent spiking neural networks, Phys Rev Lett, 118,  018103 (2017).
	     
	     \bibitem{Rosenbaum2017n}
	     R. Rosenbaum, M. A. Smith, A. Kohn, J. E. Rubin and B. Boiron, The spatial structure of correlated neuronal variability, Nature Neuroscience 20, 107–114 (2017)
	     

           \bibitem{Darshan2018}
           R. Darshan, C. van Vreeswijk, and D. Hansel,  Strength of Correlations in Strongly Recurrent Neuronal Networks, Phys. Rev. X 8, 031072 (2018).  
	       
          \bibitem{Chaudhuri2015}
          R. Chaudhuri, K. Knoblauch, M.-A. Gariel, H. Kennedy, and X.-J. Wang, A large-scale circuit mechanism for hierarchical dynamical processing in the primate cortex, Neuron 88, 419 (2015).
	       
	     
           
           \bibitem{Sompolinsky1988}
           H. Sompolinsky, A. Crisanti, and H. J. Sommers, Chaos in Random Neural Networks,
           Phys. Rev. Lett. 61, 259  (1988).
           
           
            \bibitem{Hansel2015}
            O. Harish, D. Hansel, Asynchronous rate chaos in spiking neuronal circuits, PLoS computational biology 11, e1004266 (2015).
            
            \bibitem{Kadmon2015}
            J. Kadmon and H. Sompolinsky, Transition to Chaos in Random Neuronal Networks, Phys. Rev. X 5, 041030 (2015).
            
            
          \bibitem{Meegen2021}
          A. van Meegen and S. J. van Albada, Microscopic theory of intrinsic timescales in spiking neural networks, Phys. Rev. Research 3, 043077  (2021).
          
         \bibitem{Kumar2012}
            A. Litwin-Kumar and B. Doiron, Slow dynamics and high variability in balanced cortical networks with clustered connections, Nature Neuroscience  15, 1498–1505  (2012).
	   
         \bibitem{Stern2022}
          M. Stern, N. Istrate, L. Mazzucato, A reservoir of timescales in random neural networks,  arXiv:2110.09165 (2022).    
	      
	      
	      
       \bibitem{Chaudhuri2014}
        R. Chaudhuri, A. Bernacchia, X. J. Wang, A diversity of localized timescales in network activity, elife 3, e01239 (2014).
        
	      



		\bibitem{Ginzburg1994}
		Iris Ginzburg and Haim Sompolinsky, Theory of correlations in stochastic neural networks, Phys. Rev. E 50, 3171 (1994).
		
		
		\bibitem{Renart2010}
		Alfonso Renart et al, The Asynchronous State in Cortical Circuits, Science 327 (5965), 587-590 (2010).
	
		
		\bibitem{Chow2010}
        M. A. Buice, J. D. Cowan, Carson C. Chow, Systematic Fluctuation Expansion for Neural Network Activity Equations, Neural Computation 22 (2), 377–426 (2010).
		
		
		\bibitem{Dahmen2016}
		D. Dahmen, H. Bos, and M. Helias, Correlated Fluctuations in Strongly Coupled Binary Networks Beyond Equilibrium, Phys. Rev. X 6, 031024 (2016).
		
		\bibitem{Farkhooi}
		Farzad Farkhooi and Wilhelm Stannat, Complete Mean-Field Theory for Dynamics of Binary Recurrent Networks, Phys. Rev. Lett. 119, 208301 (2017).
		


    
        \bibitem{Engel2016}
        T. A. Engel, et al., Selective modulation of cortical state during spatial attention, Science 354, 1140–1144 (2016).
            
            
        \bibitem{Ashcroft1976}
          N. W. Ashcroft and N. D. Mermin, Solid state physics, New York: Saunders College Publishing. ISBN 0-03-083993-9. OCLC 934604 (1976).
       
        \bibitem{Huang2017}
        C Huang, B Doiron, Once upon a (slow) time in the land of recurrent neuronal networks, Current opinion in neurobiology 46, 31-38 (2017)
       

        
        \bibitem{Randomband}
          Paul Bourgade,  Random band matrices,  arXiv:1807.03031v2 [math.PR] 
          
        
        \bibitem{Landau2018}  
        I. D. Landau, H. Sompolinsky, Coherent chaos in a recurrent neural network with structured connectivity, PLoS computational biology 14 (12), e1006309 (2018)
 




		
		

		

		
	
		
		
	\end{thebibliography}
\end{document}